\documentclass[fleqn,usenatbib]{mnras}

\usepackage{newtxtext,newtxmath}
\usepackage{xcolor}
\usepackage[T1]{fontenc}

\DeclareRobustCommand{\VAN}[3]{#2}
\let\VANthebibliography\thebibliography
\def\thebibliography{\DeclareRobustCommand{\VAN}[3]{##3}\VANthebibliography}

\usepackage{graphicx}	% Including figure files
\usepackage{amsmath}	% Advanced maths commands
\newcommand{\Msun}{M$_{\odot}$}

\usepackage[normalem]{ulem}
\title[Environmental Effects on Dwarf Galaxy Growth]{Large Scale Structure and Environmental Effects on Dwarf Galaxy Growth}

\author[Mac M. McMullan et al. 2026]{
Mac M. McMullan,$^{1}$\thanks{E-mail: mac.m.mcmullan@durham.ac.uk}
Sownak Bose,$^{1}$\thanks{E-mail: sownak.bose@durham.ac.uk}
Azadeh Fattahi,$^{1,2}$
Isabel Santos-Santos,$^{1}$
\newauthor
Wojciech A. Hellwing,$^{3}$
and Tilly A. Evans-Hofmann$^{4,5}$
\\
$^{1}$Institute for Computational Cosmology, Department of Physics, Durham University, Durham DH1 3LE, UK
\\
$^{2}$The Oskar Klein Centre, Department of Physics, Stockholm University, Albanova University Center, 106 91 Stockholm, Sweden
\\
$^{3}$Center for Theoretical Physics, Polish Academy of Sciences, Aleja Lotników 32/46, 02-668 Warsaw, Poland
\\
$^{4}$Physics Education Research Group, School of Physics \& Astronomy, Cardiff University, Queens Buildings, CF24 3AA, UK
\\
$^{5}$Cardiff Hub for Astrophysics Research \& Technology, School of Physics \& Astronomy, Cardiff University, Queens Buildings, CF24 3AA, UK
}

\date{Accepted XXX. Received YYY; in original form ZZZ}

\pubyear{2026}

\begin{document}
\label{firstpage}
\pagerange{\pageref{firstpage}--\pageref{lastpage}}
\maketitle

\begin{abstract}
Dwarf galaxies serve as key models for understanding galaxy assembly in the early universe, with their final properties influenced by environmental factors. Using the dark matter-only simulation “Copernicus Complexio” (COCO) and the semi-analytic model GALFORM, we examine the stellar mass assembly of dwarf galaxies across different cosmic web regions, defined by the NEXUS+/\texttt{CACTUS} algorithm. We identify significant variations in stellar mass assembly based on final mass, with the largest dwarf galaxies assembling, on average, 50\% of their mass 7.7 Gyrs later than the  smallest ones. Central galaxies also differ in their assembly from satellites of comparable final mass, forming 50\% of their mass 2.5 Gyrs later. The location within the cosmic web further influences assembly, with satellite galaxies showing greater differences than centrals. Satellites in the densest regions assemble their mass 1.5 Gyrs earlier than those in the least dense regions, compared to 0.69 Gyrs for central galaxies. This disparity arises from varying infall times, with satellites in dense environments infalling 5.2 Gyrs earlier than those in voids. Additionally, we investigate the impact of reionisation parameters, specifically the timing ($z_{cut}$) and filtering scale ($v_{cut}$) of reionisation. The stellar-to-halo-mass relation shows a power law break between $10^8$~\Msun~$< M_{200} < 10^{10}$~\Msun{}, with earlier $z_{cut}$ or higher $v_{cut}$ leading to more star formation suppression in lower-mass haloes. The halo occupation fraction is also affected, with later $z_{cut}$ or lower $v_{cut}$ resulting in fewer lower-mass haloes being occupied at $z=0$. Our investigation provides a valuable theoretical framework for interpreting upcoming observational data in this mass regime.

\end{abstract}


%%%%%%%%%%%%%%%%%%%%%%%%%%%%%%%%%%%%%%%%%%%%%%%%%%

%%%%%%%%%%%%%%%%% BODY OF PAPER %%%%%%%%%%%%%%%%%%

\section{Introduction}
\label{sec:intro}
%\WH{General notes:\\
%Consistency sweep request: please ensure terminology is uniform across the manuscript: (i) use `NEXUS+' consistently (since Sec.~2.3 defines NEXUS+ via CACTUS/Hunde2025), including in captions and Results text; (ii) define `reionisation relic' on first mention per section, then `relic' is fine; (iii) fix the `Fillament' typo in figure legends by correcting the plotting scripts and re-exporting the PDFs (LaTeX edits will not fix embedded legend text).} \Mac{These comments have now been implemented :)}

Hierarchical structure formation is the theory that the growth of dark matter haloes, and therefore galaxies, begins with small clumps, which grow to more massive galaxies through mergers as well as through smooth accretion of dark matter and gas \citep{frenk1988}. This is a vital part of our standard cosmological model, Lambda Cold Dark Matter ($\Lambda$CDM), and informs our theories of how galaxies evolve. 

The smallest galaxies observed today can give us an insight into how the galaxy population evolved from these smaller progenitors, given the evidence that many of these dwarf galaxies, especially the smallest dwarfs, have not evolved much over the last 10 billion years when compared with more massive galaxies \citep{bovill2009}. Dwarf galaxies also have large virial masses compared to their stellar masses, meaning they are less efficient at forming stars for the amount of mass they have when compared to larger galaxies \citep{white1978}. This means they can be used to constrain feedback processes \citep{benson2003, whitefrenk1991, somerville2015}, but dwarfs have also been used to probe the nature of dark matter, for example, through the core-cusp problem, which helps us to constrain the behaviour of dark matter on the smallest scales \citep{sales2022}. The number of UFDs (ultra-faint dwarfs, with stellar mass $M_\star \lesssim 10^5\,\mathrm{M}_\odot$) around the Milky Way can also help to place a limit on the mass of the dark matter particle \citep{jethwa2018, nadler2024, newton2025}%\WH{Fixed the outdated record for Newton 2024 it is now published and should be Newton 2025} \Mac{Thanks Wojciech!}

The mechanisms that regulate star formation in dwarf galaxies have attracted considerable attention, particularly in the context of tensions within the $\Lambda$CDM paradigm, such as the Missing Satellite Problem \citep{klypin1999}. This problem arises from the apparent overabundance of low-mass dark matter haloes relative to the comparatively small number of dwarf satellites observed around the Milky Way. One proposed resolution involves the suppression of star formation in low-mass systems due to the effects of reionisation \citep{rees1986, kauffmann1993, benitez-llambay2020}, in combination with other environmental and feedback processes. It is now widely accepted that reionisation strongly influences the ability of ``ultra-faint'' dwarf galaxies to form stars: once the universe is reionised at $z \approx 6$, further star formation in these systems is expected to be largely quenched. In this paradigm, ultra-faint dwarf galaxies should display relatively uniform star formation histories, with the vast majority forming the bulk of their stars prior to reionisation and have greatly suppressed star formation thereafter, becoming ``reionisation relics''. Larger dwarf galaxies are able to re-start star formation post-reionisation through growing their dark matter haloes, or avoid the effect of reionisation completely, leading to a stark divide between reionisation relics and normal dwarf galaxies, where the normal dwarf galaxies continue to grow, while the relics remain near the stellar mass they attained at reionisation. Such a bimodality is expected to leave a distinct imprint on the stellar mass function of galaxies \citep{bose2018}. However, observations suggest a more complex scenario: some ultra-faint dwarfs exhibit renewed bursts of star formation even after reionisation is thought to have ended at $z \sim 6$ \citep{mcquinn2023pegW}.

The interest in the star formation histories of dwarfs has grown as more observational data has become available. Notably, a diversity of star formation histories are observed in dwarf galaxies, and this is also true in the ultra-faint regime. \citet{weisz2014} re-constructed the star formation of 40 Local Group Dwarf Galaxies and found that there was a large diversity in quenching times, even among ``ultra-faint'' dwarfs (UFDs). They discovered that the observed fraction of stellar mass formed before $z=2$ was 80\% for galaxies with $M_\star < 10^5$ M$_{\odot}$ and 30\% for galaxies with $M_\star > 10^7$ M$_{\odot}$, and that this difference was likely due to a difference in the times at which the galaxies became satellites, or whether the UFDs were embedded in the haloes of the Andromeda galaxy (M31) or the Milky Way. \citet{weisz2019} later found that M31 satellites do not show a similar trend of fainter galaxies quenching earlier, and instead have possibly been more strongly influenced by a recent merger in Andromeda's history. The Milky Way satellites, by comparison, follow a relation that links the faintness of the satellite with an earlier quenching time, and the differences in infall time of the satellites being the explanation for diversity between objects of similar brightness. This indicates that the accretion history of the host can influence the quenching times of the satellites, but also indicates that the conclusions we draw from the quenching times of satellites of more isolated host galaxies may not apply to galaxies with more active merger histories.

Recently, several more very low mass galaxies have been discovered in the Local Group, and whose star formation histories have been mapped. In \citet{mcquinn2023pegW,mcquinn2024P,mcquinn2024MK}, the authors find four new ultra-faint dwarf galaxies within the Local Group, but outside the virial radii of the Milky Way and Andromeda. For Leo M and Leo K, characterised in \citet{mcquinn2024MK}, the authors find that the galaxies form the majority of their stars before and during the epoch of reionisation, whereas for Leo P and Pegasus W, the authors infer a star formation history in which the majority of star formation occurred post-reionisation. This could mean that we are entering a regime observationally in which we can start to constrain the physics of the reionisation process itself, at least locally, using the stellar mass assemblies of these isolated ultra-faint galaxies.

Yet another factor that could influence the assembly of dwarf galaxies is the large-scale environment in which they reside. \citet{thomas2010} find that the impact of changing environment on galaxies  increases with decreasing galaxy mass or, in other words, those with the shallowest potential wells. %Shallow potential wells mean the galaxy is much more sensitive to internal feedback mechanisms and external forces such as reionisation due to low gravity. 
Recent research has found that the majority of ionising photons in the universe originated in dwarf galaxies because they can easily escape their shallow potential wells, as well as the authors discovering that the efficiency at which dwarfs produce ionizing photons is higher than was previously thought \citep{atek2024}. If these dwarf galaxies are the main contributors to reionisation, this could indicate that proximity to a region with more massive galaxies should not affect dwarf galaxy reionisation time, because massive galaxies are not the main drivers of reionisation. However, if, as explored in \citet{naidu2020}, the most massive galaxies are the drivers of reionisation, a dwarf galaxy's local environment would have a very noticeable effect on its star formation history, with those close to large galaxies ionizing faster than those in the field.

%ully understand dwarf galaxies in the context of our current model of galaxy formation, we must use simulations to make predictions using our current models of galaxy formation, which can then be compared to data from future surveys. For example, in \citet{davis1985}, the authors used a simulation of their model of cold dark matter to predict what the large scale structure of the universe would look like, and were proven right 17 years later when the 2MASS survey showed that the distribution of galaxies did follow the predictions made in 1985 \citep{norberg2002}. Another example of using simulations to make predictions for our models is found in \citet{benson2000}, in which GALFORM is used to ``solve" the aforementioned ``Missing Satellite Problem", many years before it was considered to be solved by the wider community.}

Cosmological simulations provide a powerful framework for studying the formation and evolution of dwarf galaxies. Hydrodynamic simulations follow both baryonic and dark matter components by directly solving the equations of gravity and gas dynamics, allowing processes such as gas accretion, cooling, and feedback to emerge self-consistently. Semi-analytic models (SAMs)  describe galaxy formation processes through sophisticated parameterised prescriptions applied to halo merger trees. The major benefit of SAMs is that they are computationally far cheaper to run, meaning that they can be run repeatedly and parameters or galaxy evolution processes as a whole can be changed with relative ease. SAMs can be run on Monte Carlo merger trees, or on merger trees from DM-only simulations. For example, \citet{yayura2023} use the Semi-Analytic Galaxies model (\textsc{SAG}) applied to the Small MultiDark PLank simulation (SMDPL) to study the stellar-halo mass relation in different environments, while \citet{monzon2024} employ \textsc{SatGen} to explore scatter in this relation around a Milky Way analogue using Monte Carlo merger trees.

When using a simulation with N-body dynamics of any kind, a common strategy is the ``zoom-in'' technique, in which a region within a larger cosmological volume is re-simulated at a higher resolution. This technique can be used to identify specific environments in a low-resolution volume to study at a higher resolution. Examples include the Latte project \citep{wetzel2016} and DC-Justice League \citep{christiansen2024}, which simulate Milky Way-mass galaxies and their dwarf satellites, as well as the Romulus suite, which targets both galaxy clusters \citep{tremmel2019, tremmel2020} and isolated field dwarfs \citep{tremmel2017}. Field dwarfs are also the focus of the MARVEL-ous Dwarfs simulations \citep{christiansen2024}. High-resolution zoom-ins of a handful of haloes \citep[e.g.][]{agertz2020, gutcke2021} can model dwarf galaxies with great physical fidelity, but lack the ability to probe large populations or situate dwarfs within their full large-scale structure (LSS) context.

%\WH{Insert here:  add a brief state-of-the-art paragraph on cosmic-web classifiers, to (i) acknowledge other common approaches, and (ii) justify why a cosmic-web (not just local-density) environmental split is meaningful. Cite a compact but representative set: Libeskind et al. 2018 review/benchmark; Forero-Romero 2009 (tidal tensor); Hoffman et al. 2012 (velocity shear); Sousbie 2011 (DisPerSE); Aragón-Calvo et al. 2007 (MMF lineage); Tempel et al. 2014 (Bisous filaments); Kraljic et al. 2018 (GAMA + DisPerSE distances).}

%% note below paragraphs written by Wojciech, thanks Wojciech!
A complementary and more geometric notion of environment is provided by the cosmic web, commonly decomposed into voids, walls (sheets), filaments, and nodes. A wide range of operational classifiers exists, and different methods emphasise different physical or topological aspects of large-scale structure \citep[e.g.][]{libeskind2018}. Popular families include deformation- or tidal-tensor schemes \citep[e.g.][]{forero2009}, velocity-shear based classifications \citep[e.g.][]{hoffman2012}, and topological skeleton approaches such as DisPerSE \citep{Sousbie2011}, which are often used to connect galaxy properties to distances from filaments, walls, and nodes \citep[e.g.][]{kraljic2018}. Other widely used approaches include multiscale morphology filters \citep[e.g.][]{aragoncalvo2007} and stochastic filament finders such as the Bisous model \citep[e.g.][]{tempel2014}. These complementary perspectives motivate explicitly testing whether dwarf-galaxy assembly varies across cosmic-web environments, beyond trends with halo mass and the central-satellite split. Importantly, Hessian-based cosmic-web classifications often involve a thresholding choice that can affect inferred volume fractions and environmental trends, motivating physically motivated prescriptions for more robust and comparable results \citep{Olex2025}.

%\WH{Add a short state-of-the-art paragraph on empirical and modelling results linking galaxy properties to cosmic-web environment, to motivate why a cosmic-web split is scientifically informative (not just a classification exercise), and why dwarfs are a sensitive regime for such trends. Please ensure the citation set includes at least one observational anchor and one modelling/SAM anchor, in addition to our key references (Hellwing+2021; Jaber+2024).}

Using these classifiers, a growing literature has investigated how galaxy properties vary across cosmic-web environments or as a function of distance to filaments and nodes. Observational analyses report systematic gradients in stellar mass and star-formation activity relative to filament spines and node regions, suggesting that the cosmic web can modulate galaxy growth beyond what is captured by a purely local-density description \citep[e.g.][]{kraljic2018,malavasi2017,alpaslan2016}. Complementary modelling work, including studies based on semi-analytic galaxy catalogues, finds that a cosmic-web split can highlight differences in assembly histories and accretion pathways, with particularly clear signatures in lower-mass systems where satellite infall times and environmental processing play an outsized role \citep[e.g.][]{Jaber2024,Hellwing2021}. This motivates testing whether the diversity of dwarf-galaxy assembly histories can be partially attributed to their location within the cosmic web, and whether a cosmic-web split provides information beyond the standard central--satellite dichotomy.
%% above paragraphs written by Wojciech

%\WH{Intro: suggest adding a short paragraph pointing out that COCO has already been used in several related studies of very faint and low-surface-brightness dwarf-galaxy populations (e.g. dwarf stellar haloes, particle-tagged accreted components, and ghost-galaxy/ultra-diffuse scenarios). This helps motivate why COCO is an appropriate platform for connecting dwarf assembly diversity with environment and with upcoming surveys, without making this paper a methods advert.}
The aim of this paper is to investigate the extent to which environment could be responsible for the diversity of dwarf galaxy star formation histories. The behaviour of dwarfs within different large-scale environments has been difficult to investigate in simulations due to the competing need for a large volume simulation with high resolution. In our work, we utilise the  ``Copernicus Complexio'' (COCO) simulations from \citet{hellwing2016}, which is a high resolution zoom-in region of the dark matter-only COLOR (``COpernicus complexio LOw Resolution'') simulation. The dark matter particles are $1.6  \times 10^5$\Msun, allowing us to resolve dark matter haloes down to $3.2 \times 10^6$\Msun, and therefore also resolve all the galaxies that could form in haloes above the atomic cooling limit. We then apply the semi-analytic model GALFORM \citep{lacey2016} to this DM-only simulation to output the properties of galaxies residing within this volume. This allows us to resolve even the smallest dwarfs within a large cosmological environment with relatively little computational expense. 
The COCO simulation has also been used in a series of complementary studies aimed at interpreting extremely low surface-brightness dwarf-galaxy populations and faint stellar structures, including the expected diversity of accreted stellar haloes in low-mass field galaxies when combining COCO with semi-analytic modelling and particle-tagging techniques \citep[e.g.][]{Cooper2025Accreted,Deason2022DwarfHalos,MiroCarretero2025Streams}. Related work has explored `ghost galaxy' formation channels in which low-mass haloes acquire most of their stellar mass through mergers rather than in-situ star formation, producing diffuse, accretion-dominated systems that may be detectable in deep wide-area imaging \citep{Wang2023Ghostly}.

\section{Theory}
\label{sec:theory}

\begin{figure*}
	\includegraphics[width=16cm]{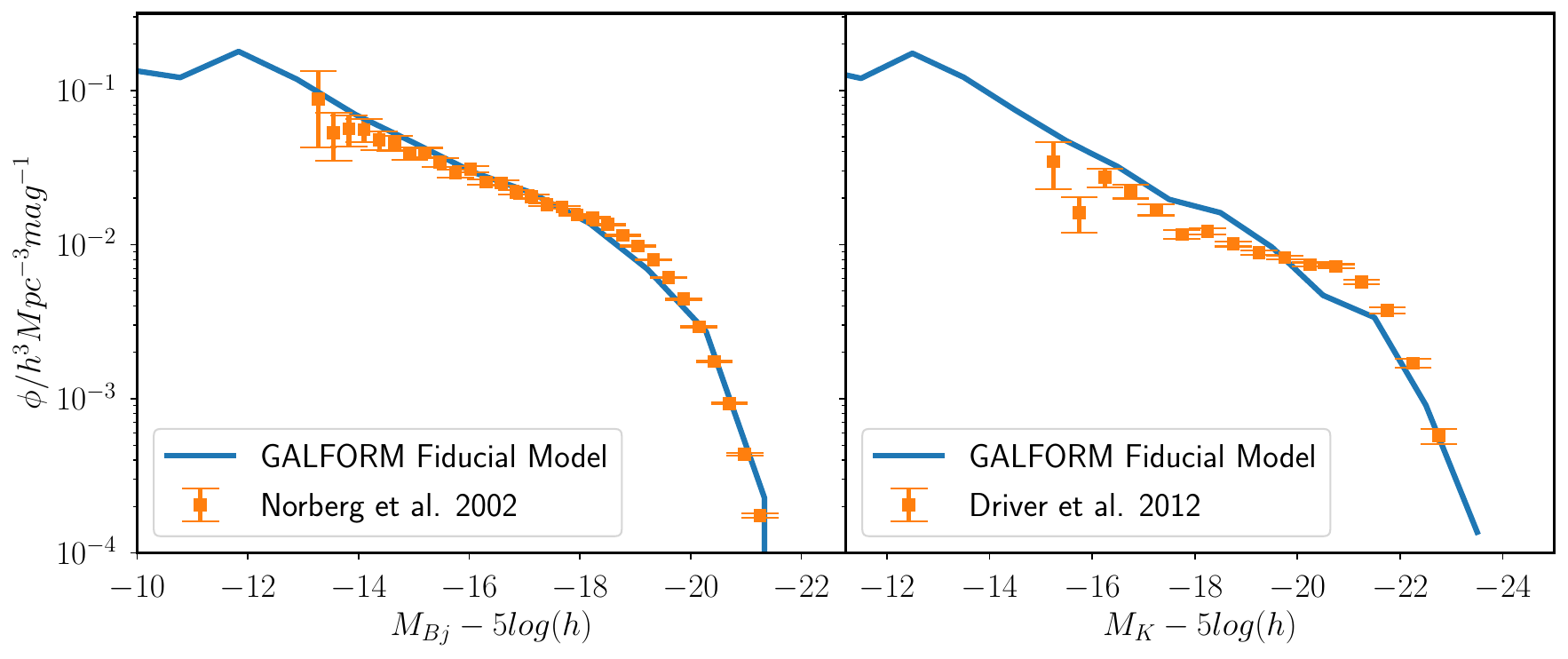}
    \caption{The luminosity function at $z=0$ for all galaxies in the fiducial model matched with the $b_J$-band and $K$-band luminosity functions from \citet{norberg2002} and \citet{driver2012}, respectively. Parameters in GALFORM are tuned to match these luminosity functions by adjusting a few parameters in the model by a small amount, in the regime where data is available, detailed in \citet{lacey2016}.}
    \label{fig:LF_both}
\end{figure*}

\begin{figure}
	\includegraphics[width=\columnwidth]{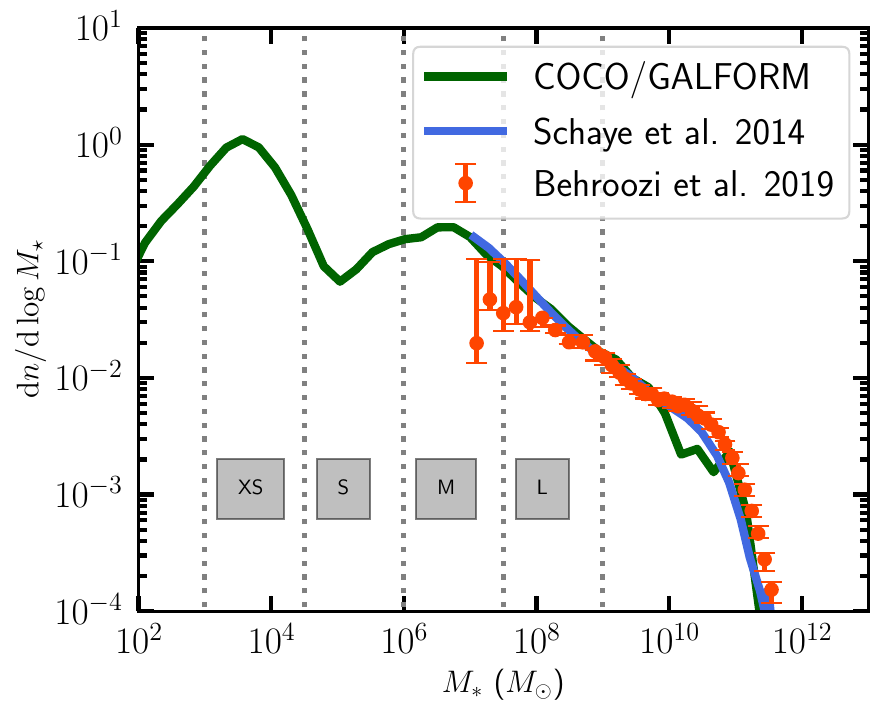}
    \caption{The stellar mass function of all galaxies in the COCO/GALFORM fiducial model, shown compared with the EAGLE simulation \citep{schaye_2015}, and observational data compiled by \citet{behroozi2019}, which includes data from \citet{moustakas2013} and \citet{baldry2012}. There is a good agreement between GALFORM and EAGLE, and reasonable consistency with the observational data. Note, however, a detailed comparison with the inferred stellar mass function at $z=0$ is not straightforward. This is because GALFORM uses two different initial mass functions (IMFs), one for starbursts and one for quiescent star formation, whereas a single IMF is typically assumed in observations. The nature of this SMF is bimodal, with the two populations being the reionisation relics and the larger galaxies. See Section~\ref{sec:mass_ass} for more information on these groups. The dip in the SMF corresponds to the mass scale where reionisation affects galaxy formation in the model. The labels XS, S, M and L correspond to the mass bins used later in Section~\ref{sec:fiducial_results}.}
    \label{fig:SMF_fiducial}
\end{figure}

\begin{figure}
	\includegraphics[width=\columnwidth]{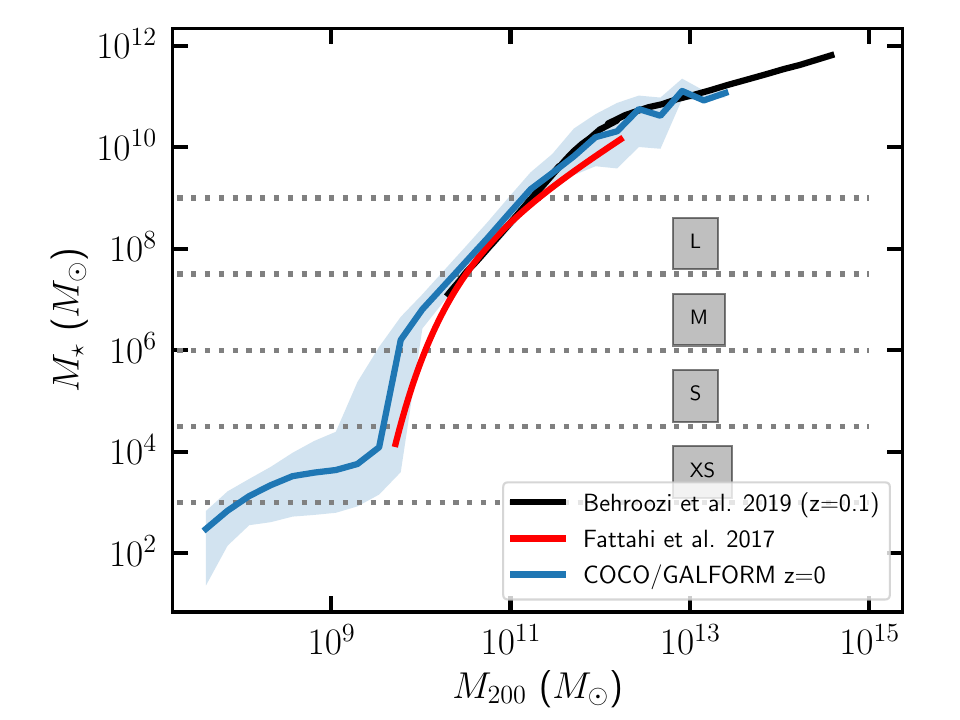}
    \caption{The $z=0$ stellar-to-halo-mass relation for central galaxies in the fiducial model used in this paper, compared to the relations from \citet{fattahi2017} and \citet{behroozi2019}. The blue line is the 50th percentile of the COCO/GALFORM data, with the shaded region representing the 16th to 84th percentiles. The relation is very steep from $10^{4} - 10^{6} $\Msun, meaning that galaxies with vast differences in stellar mass can be formed within haloes of very similar final masses, as is also seen in Figure~\ref{fig:RI_halogrowth}. Both the relations from \citet{behroozi2019} and from \citet{fattahi2017} agree well with the predictions from COCO/GALFORM, showing reasonable agreement between semi-analytics, empirical, and hydrodynamical simulations.}
    \label{fig:mstars_mhalo_fiducial}
\end{figure}

\begin{figure}
	\includegraphics[width=\columnwidth]{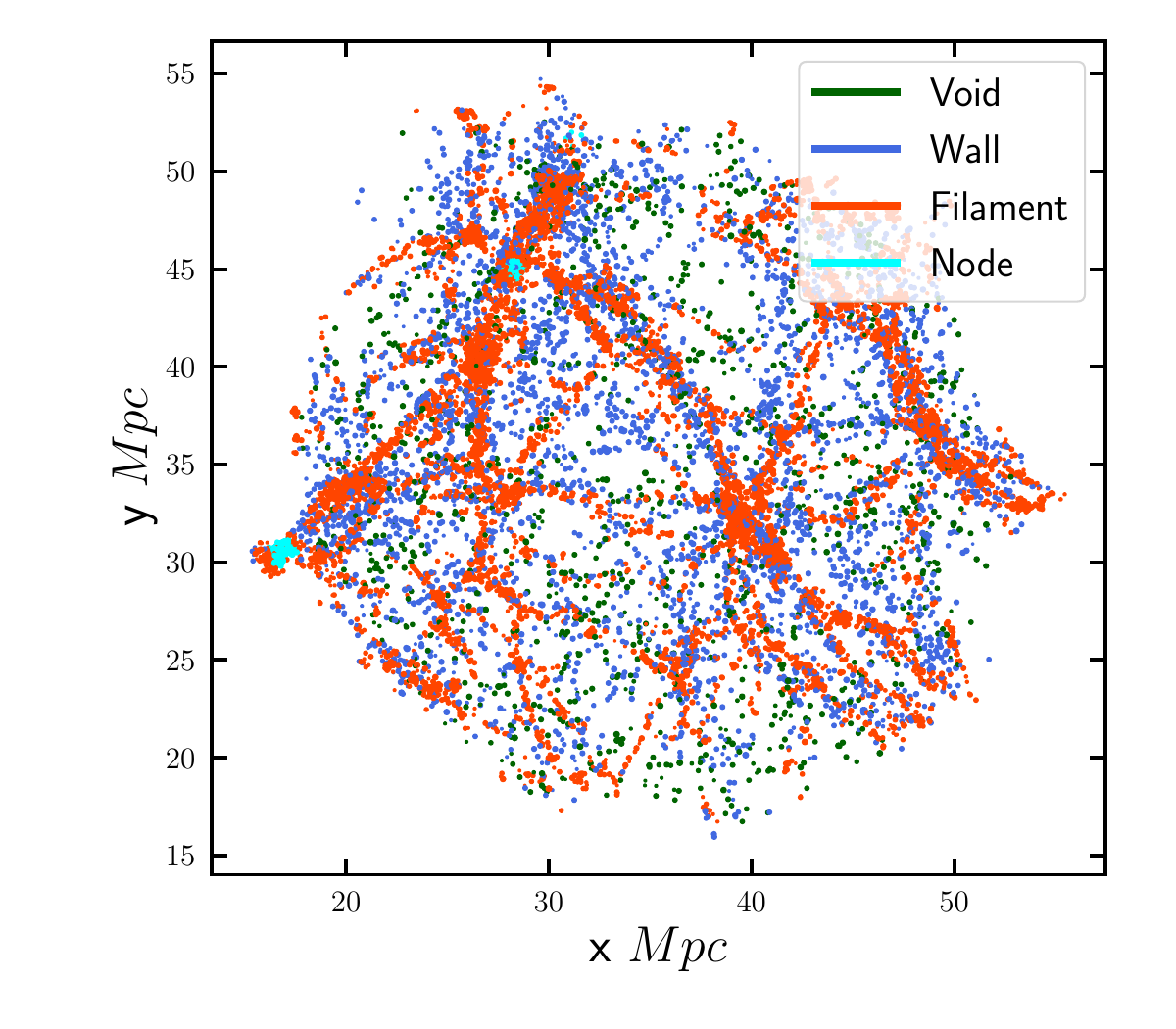}
    \caption{The COCO volume with different colours denoting the different regions of the cosmic web, as defined by NEXUS+/\texttt{CACTUS}. This is a position plot in the x-y plane, and all points have $z$ coordinates between 30 and 40. The size of the points denoting each galaxy are proportional to the stellar mass of the galaxy, using a log scale. The cosmic web structure is clearly visible.}
    \label{fig:coco_pic}
\end{figure}

In this section we introduce the dark matter simulation, COCO, and the semi-analytic model, GALFORM, which we use to investigate the link between environment and mass assembly in dwarf galaxies. The code used to characterise the large-scale structure, NEXUS+, will also be discussed.

\subsection{Copernicus Complexio (COCO)}
\label{sec:coco}
COCO is a dark matter-only zoom-in volume in the larger simulation COLOR (Copernicus Complexio Low Resolution) \citep{hellwing2016}. COLOR has a volume of $\left(100 \, \mathrm{Mpc}\right)^3$ and COCO is a roughly spherical region in the centre, measuring $68$ Mpc in diameter. The dark matter particles in the high-resolution region have a mass of $m_p = 1.6 \times 10^5 $\Msun, and the minimum resolvable halo mass is $3.2 \times 10^6 $\Msun, equivalent to $20$ high-resolution particles. This means we are able to resolve galaxies with halo masses below $10^8 $\Msun{} comfortably, although all galaxies in our model form in haloes larger than this.%, as shown in Figure~\ref{fig:occfrac_diffmod}. 

COCO was run using the GADGET-3 Tree-PM N-body code, an updated version of the publicly available GADGET-2 code \citep{springel2005}, with substructures identified using the SUBFIND algorithm \citep{springel2001}. Initial conditions were set at $z=127$ and the region is evolved up to $z=0$. The simulation adopts the cosmological parameters determined by the seven-year Wilkinson Microwave Anisotropy Probe (WMAP7) results \citep{komatsu2011}. Specifically, we assume a matter density parameter of $\Omega_{\mathrm{m}} = 0.272$ and a cosmological constant of $\Omega_{\Lambda} = 0.728$. The dimensionless Hubble parameter is taken to be $h = 0.704$, which corresponds to a Hubble constant of $H_{0} = 100\,h\,\mathrm{km\,s^{-1}\,Mpc^{-1}}$. The primordial power spectrum is assumed to be a power law with a scalar spectral index $n_{s} = 0.967$. The linear matter power spectrum is normalised at $z=0$ such that $\sigma_{8} = 0.81$. %GADGET-3 uses a hybrid method of calculating forces, with the longer range forces calculated using a particle mesh (PM) method, and the shorter-range forces calculated using a tree method. This allows the code to calculate the longer-range forces with less computational power, and focus the attention on the shorter-range forces which require the most computational power. The computing resources are therefore used conservatively, while not compromising on the results for smaller regions. The COCO simulation allows us to resolve the host haloes of ultra-faint galaxies in a large simulation volume, which will enable us to investigate the effect of the large scale structure on the star formation histories of dwarfs.

\subsection{GALFORM}
\label{sec:galform}
The GALFORM semi-analytic model \citep{lacey2016} is used in this paper to provide us with predictions for the properties of galaxies that form within dark matter haloes in COCO. The semi-analytic philosophy is best described as an effort to understand the key physical processes which contribute to galaxy evolution, and to implement these to the best of our understanding in order to reproduce the observed properties of galaxies in the real universe. This is achieved using a physically-motivated framework with parameters that can be tuned (within some reasonable range), and whose values are determined by matching to various relations at present-day. In GALFORM, this is primarily the luminosity functions in the K-band and b$_J$ band at $z=0$, as seen in Figure~\ref{fig:LF_both}. The resulting code is made up of sections governing physical processes we observe in galaxies (e.g. gas cooling, star formation, feedback from supernovae and black holes etc.). The flexibility and computational efficiency of semi-analytic models can be used to readily explore areas of galaxy formation where our understanding is lacking, and quickly add other physical processes or change parameters to experiment with alternative theories. 

The original model was based on work done by \citet{white1991}, and has been refined since its first iteration in \citet{cole2000}. Many versions of GALFORM have emerged since then to match the increasing data and knowledge we have about galaxy formation processes. For example, \citet{baugh2006} introduced a top-heavy IMF in starbursts to reproduce the observed number count of submillimetre galaxies, and \citet{bower2006} added a model of Active Galactic Nuclei (AGN) feedback to reproduce the exponential tail at the bright end of the galaxy luminosity function. Later additions include the addition of molecular gas content affecting star formation in \citet{lagos2011}, black hole spin in \citet{griffin2019} and a more complex gas cooling method in \citet{hou2018}. \citet{lacey2016} is an update to the original GALFORM from \citet{cole1994}, including AGN feedback and other additions to the original model.

The advantage of using GALFORM to study populations of dwarf galaxies is that we are able to resolve smaller galaxies within a larger simulation volume compared to hydrodynamic simulations. Because of these larger volumes, we can justify that the GALFORM model is predicting a realistic population of galaxies by comparing to data that exists for the larger galaxies, and use these comparisons to validate our model. Utilising this method, we can say that the dwarf galaxies are genuine predictions of our model. Another advantage of the GALFORM model is its speed. For this project, we have run GALFORM many times with different parameters, which would not be possible with a hydrodynamic simulation.

%The \citet{lacey2016} model includes feedback from AGN and supernovae, stellar winds, star formation in disks of galaxies, starbursts, metallicity evolution of the gas, chemical evolution of stellar populations, and importantly for this project, the cooling of gas into dark matter haloes, which includes a simple model of reionisation.

Gas in GALFORM is considered to cycle between cold gas within the galaxy, hot gas within the galaxy, and a hot gas reservoir outside the virial radius of the galaxy. Cold gas is available for star formation and to feed the central black hole, both processes then cause the gas to be released as hot gas and cool back down within a cooling timescale. A galaxy receiving a continuous supply of cold gas can continue to form stars, but galaxies can lose their access to cold gas as a result of the gas being prevented from cooling from the reservoir. When a central galaxy becomes a satellite of a larger galaxy in GALFORM, the satellite then slowly uses up its cold gas reserves and new gas is prevented from cooling onto the galaxy. Reionisation in GALFORM affects the cooling of gas into haloes in a similar way: if the halo circular velocity is below a certain value, $v_{\mathrm{cut}}$ ($ = 30{}$km{}~s$^{-1}$ in our fiducial model), below redshift $z_{\mathrm{cut}}$ ($= 6$ in our fiducial model), gas is prevented from cooling into that halo further.

The star formation rate (SFR) in galaxy discs in GALFORM is governed by:
\begin{equation}
    \Psi_{\mathrm{disc}} = \nu_{\mathrm{SF}} f_{\mathrm{mol}} M_{\mathrm{cold, disc}},
	\label{eq:quadratic}
\end{equation}
where $\nu_{\mathrm{SF}}$ is an adjustable parameter for $1\sigma$ scatter around a value of $0.43 $Gyr$^{-1}$, $f_{\mathrm{mol}}$ is the fraction of cold gas which is in a molecular form, as opposed to atomic cold gas, and $M_{\mathrm{cold, disc}}$ is the mass of the cold gas in the disc component of the galaxy. Star formation is therefore proportional to the amount of cold gas in the galaxy. For a more detailed explanation of star formation in GALFORM, see section $3.4$ in \citet{lacey2016}. 

The GALFORM model is very successful at reproducing relationships found in observations of galaxies, including luminosity functions, black hole-bulge mass, and morphological relations. In Figure~\ref{fig:LF_both}, we show the luminosity function in the $b_{J}$ and $K$-bands, which are used to calibrate the parameters of the GALFORM model. The luminosities of the galaxies in GALFORM are found  using the stellar masses of the galaxies, combined with a stellar population synthesis model \citep{maraston2005}. We then use a simplified two-temperature model for dust attenuation. After this, GALFORM finds the broadband luminosities and magnitudes by convolving the predicted stellar spectral energy distributions with filter response functions for various observational bands. Our philosophy in GALFORM is to calibrate directly to the observed light (i.e. the galaxy luminosity function) rather to the stellar masses of galaxies, as it is only the former that is observed directly. GALFORM has also been successful at reproducing the satellite luminosity function \citep{bose2018}, which can assure us that the \citet{lacey2016} model can also be used to look specifically at satellite and dwarf galaxies, having already had a degree of success in this regime.
%This is the fundamental backbone of any semi-analytic model, as it is the best measure of galaxy qualities without having to measure the masses of the galaxies involved, and therefore be subject to those observational assumptions.}

The stellar mass function (Figure~\ref{fig:SMF_fiducial}) shows a good, but not perfect agreement with observational data. This is because the \citet{lacey2016} model uses two different IMFs to generate stars: an x=0.4 \citet{kennicutt1983} IMF for quiescent star formation, and a top-heavy x=1 IMF for starbursts. When stellar mass functions are inferred from observations, this is done by fitting stellar population models to galaxy spectral energy distributions (SEDs), which requires one to assume a single IMF. Because GALFORM uses multiple IMFs, any comparison to observations is not fully consistent. In addition to this, stellar population models differ in which stellar masses fit to which SEDs. In this case, we see it as more beneficial to compare our data with the luminosity function, as this can be known directly from observations, without assuming anything about the IMFs of the galaxies. Full details on this can be found in \citet{lacey2016}. There is a characteristic dip in the stellar mass function at $M_{\star} = 10^5 $\Msun. This is due to our choices of parameters for reionisation \citep{bose2018}. The population of galaxies in GALFORM below $M_{\star} = 10^5 $\Msun{} are referred to as reionisation relics, meaning their star formation is stopped by reionisation and does not restart, whereas the population of galaxies above this mass are not totally quenched by reionisation, and have some time after reionisation with which to form stars (see Figure~\ref{fig:RI_halogrowth}).

In Figure~\ref{fig:mstars_mhalo_fiducial}, we show the stellar-to-halo-mass relation for central galaxies predicted by GALFORM, compared to the relation from the APOSTLE simulations \citep{fattahi2017} and \citet{behroozi2019}. The stellar-to-halo-mass relation in dwarfs is driven by the different populations of galaxies within the dwarf regime. The relation is extremely sensitive to reionisation, whose tell-tale sign is a break from a power law at a mass scale of $\sim10^9\,\mathrm{M}_\odot$ in halo mass. %as the stellar masses of the different dwarf populations (discussed in more detail in Section~\ref{sec:mass_ass}, and directly shown in Figure~\ref{fig:mstars_mhalo_diffmod}) are defined by reionisation. 
The shallow region of Figure~\ref{fig:mstars_mhalo_fiducial} between $M_{200} = 10^{7.8} - 10^{9.5} $\Msun, corresponding roughly to the ``XS'' mass bin, is a region where star formation is extremely suppressed by reionisation. The galaxies slightly larger, in the stellar mass regime between $M_{\star} = 10^{4.5} - 10^6 $\Msun, corresponding to the ``S'' mass bin (the steep region of Figure~\ref{fig:mstars_mhalo_fiducial}), are affected by reionisation but not as completely. The results shown in section~\ref{sec:theory} and section~\ref{sec:fiducial_results}, use the reionisation parameters of our fiducial model, where reionisation happens at $z = 6$, and the circular velocity of the haloes affected is 30 km{}~s$^{-1}$ and below.

Galaxies are assigned ``types'' within GALFORM, corresponding to their relationship to galaxies around them. For central galaxies, the galaxy must be the largest galaxy within its halo, and gas cools onto central galaxies as normal. Satellites are galaxies whose haloes have become embedded into the halo of a larger central galaxy. As mentioned previously, gas in \citet{lacey2016} does not cool onto satellites. These galaxies then use up their remaining cool gas, and eventually stop star formation. (There are more recent versions of GALFORM which have experimented with allowing gas to cool onto satellites, namely \citet{hou2018}, with mixed success.) Satellite galaxy dark matter haloes are also considered to stop evolving at the moment of infall because the haloes are subject to disruption within the host halo, and it is difficult for SUBFIND to estimate what their mass is. The final galaxy type is orphan galaxies. These galaxies  formed within dark matter haloes that were, at one point, above the resolution limit of COCO, but after becoming satellites, lose mass through tidal stripping of DM particles before falling below the minimum resolution limit of 20 particles defined by the SUBFIND halo finder. At this point, GALFORM keeps the final mass from when the orphan was still resolved, and COCO tracks the most-bound particle. Therefore, once a galaxy becomes an orphan, it cannot evolve further within GALFORM, unless it merges onto a galaxy above the resolution limit. Orphan galaxies are dominant in the ultra-faint regime, as is shown in Figure~\ref{fig:galtype_frac}, and are often missed out in other simulations, despite their inclusion being necessary to reproduce the observed distribution of satellites orbiting close to the Milky Way \citep{santos-santos2024}. Note that we do not include the impact of stellar mass loss through tidal stripping once a galaxy becomes a satellite or an orphan; the stellar mass quoted therefore corresponds to their {\it peak} stellar mass.

\subsection{NEXUS+}
\label{sec:Nexus}
%\WH{Sec.~2.3 (NEXUS+): the current description reads like a tidal-tensor / potential-Hessian (T-web) classifier. Please revise to describe the NEXUS/NEXUS+ workflow as used here, i.e. NEXUS+ flags computed with the \texttt{CACTUS} implementation (per agreement cited via Hunde et al.~\citep{Hunde2025}), applied to the density field with multiscale filtering. Also make explicit the operational definition used in this paper: environment label from the grid cell containing the halo centre, and classification applied at $z=0$.} \Mac{Thanks for the work on this section, very helpful!}

% below paragraph added by Wojciech
NEXUS, introduced in \citet{cautun2013}, is a cosmic-web classification algorithm applied to a gridded representation of the matter density field, which identifies four environments, voids, walls, filaments and nodes (listed here in order of increasing density). In this work we use the NEXUS+ implementation provided within the \texttt{CACTUS} pipeline \citep{Hunde2025}, and we adopt the corresponding $z=0$ cosmic-web flags for the COCO volume (see more details in \citealt{Hellwing2021} for the first introduction of the NEXUS+ segmentation of COCO used in this context). NEXUS+ is a multiscale, Hessian-based morphological filter, in which the environmental signature is evaluated across a set of smoothing scales and combined into a scale-independent classification. Haloes (and their galaxies) are assigned an environment label using the flag of the grid cell containing the halo centre. As running the full NEXUS+ pipeline is computationally expensive, we apply the cosmic-web classification only to the $z=0$ output of COCO. A slice of the volume with the NEXUS+ flags applied to the galaxies is shown in Figure~\ref{fig:coco_pic}.

\section{Results}
\subsection{Fiducial Model}
\label{sec:fiducial_results}
The fiducial model, which was the primary model during our investigation, assumes a $z_{\mathrm{cut}} = 6$ and $v_{\mathrm{cut}}= 30$ km{}~s$^{-1}$. In the context of GALFORM, $z_{\mathrm{cut}}$ is best understood as the time at which reionisation is over, and $v_{\mathrm{cut}}$ dictates which haloes are affected by reionisation. Our choices of $z_{\mathrm{cut}}$ and $v_{\mathrm{cut}}$ are based on the model in \citep{bose2018}, and the justifications within, which reproduces the satellite luminosity function of the Milky Way. This is desirable given our focus on low mass and satellite galaxies. Other models, which vary both $z_{cut}$ and $v_{cut}$  are examined in Section~\ref{sec:alteration}.

\subsubsection{Stellar Mass Assembly for Different Mass Scales}
\label{sec:mass_ass}
\begin{figure}
	\includegraphics[width=8cm]{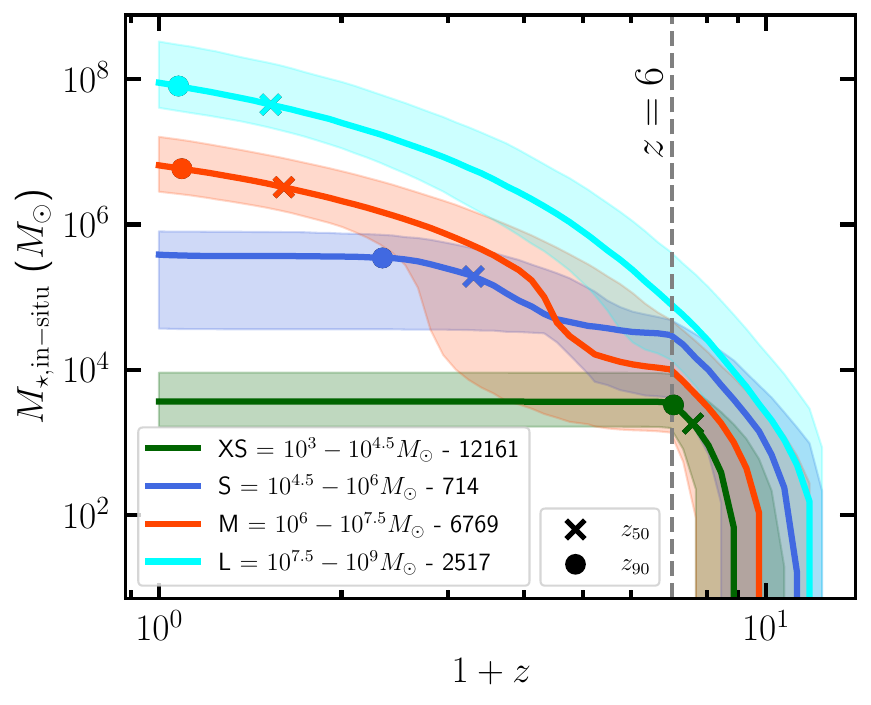}
    \caption{The in-situ stellar mass assembly of central dwarf galaxies in COCO/GALFORM, here split into four different stellar mass bins with a log y-axis. The points show the $z_{50}/m_{50}$ and $z_{90}/m_{90}$ points, marking the points at which the stellar mass of the galaxy is at 50\% and 90\% of its final mass respectively, and the shaded regions represent the 16th to the 84th percentile of assembly at a given redshift. The final day mass of an object has a significant impact on the stellar mass assembly, with the smallest galaxies assembling the majority of their stellar mass before  $z=6$, and the largest objects assembling most of their stellar mass after $z=2$. The ``S'' mass bin are ``turnover dwarfs'', with masses between lower-mass reionisation relics and ``normal'' larger galaxies. The largest mass bin, ``L'' assembles 50\% of its mass at a redshift ($z_{50}$) 7.7 Gyrs later (on average) than the smallest mass bin, ``XS''. The numbers next to each mass bin in the legend represent the number of galaxies in each mass bin.}
    \label{fig:mass_SMA}
\end{figure}

Traditional divisions within the dwarf galaxies mass range are based on observational effects, such as the brightness of the objects. In \citet{bullock2017}, dwarfs are divided into $3$ catagories: bright dwarfs, which are at the limit of completeness for faint galaxies in field surveys, and have masses between $M_{\star} = 10^7 - 10^9 $\Msun; classical dwarfs, which were the faintest objects known prior to the Sloan Digital Sky Survey (SDSS), and have masses in the range $M_{\star} = 10^5-10^7 $\Msun; and ultra-faint dwarfs, which they define as being so dim as to only be discovered after the advent of the SDSS, and which have masses below $M_{\star} =10^5 $\Msun.

The in-situ stellar mass assembly, which represents the stellar mass of a galaxy formed within the main branch of the halo tree by some redshift, $z$, can be interpreted as a measure of the growth of a galaxy. In Figure~\ref{fig:mass_SMA}, we can see the average cumulative stellar mass assembly of all galaxies, divided into mass bins based on their $z=0$ in-situ stellar mass. Lower-mass dwarf galaxies, in general, tend to form the bulk of their stars at $z>2$, whereas  higher-mass galaxies in the dwarf range continue forming stars long after this time. 

\begin{figure}
	\includegraphics[width=8cm]{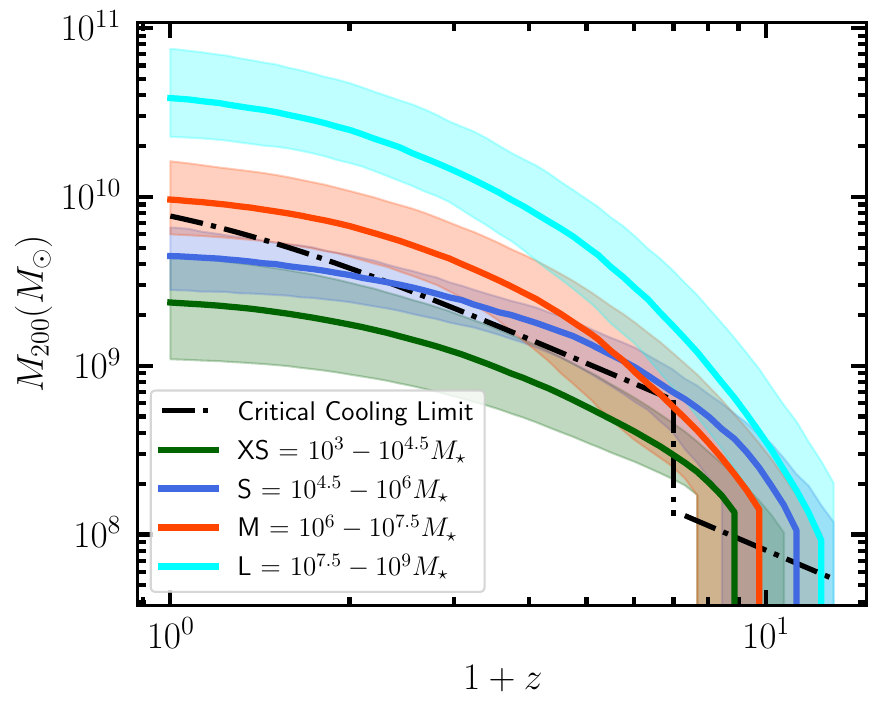}
    \caption{The growth of central dark matter haloes for the different stellar mass bins, and their interaction with our reionisation parameters. The dashed line is the halo mass limit for cooling gas into haloes imposed by reionisation after $z=6$, and before $z = 6$ is the atomic gas cooling limit. Below the line, haloes are considered too small to cool gas for star formation. Above the line, haloes are large enough to cool new gas into their galaxies to form stars. The solid coloured lines represent the 50th percentile growth within that mass bin, and the shaded areas cover from the 16th to the 84th percentile. The mass bins listed in the legend are stellar mass bins, and the values on the y-axis are the halo masses of those galaxies. All galaxies in this plot are also centrals.}
    \label{fig:RI_halogrowth}
\end{figure}

Dwarf galaxy stellar mass assembly can be roughly grouped into 3 characteristics, most influenced by the mass of the galaxy in question:
\begin{enumerate}
    \item Reionisation relic dwarfs, which are the smallest dwarfs. These formed the vast majority of their mass before reionisation (in this simulation, the fiducial model has $z_{cut} = 6$). Due to the size of their haloes, they cannot cool new gas into their centres after $z = 6$, because their circular velocity is lower than the $v_{cut}$ specified in our GALFORM run. These dwarfs can be seen in Figure~\ref{fig:mass_SMA} in the stellar mass bin XS =$10^3 - 10^{4.5} $\Msun.
    \item Larger mass dwarfs which, similar to more massive galaxies, form most of their stars after $z=2$. They are not strongly affected by reionisation because their halo circular velocity is larger than the $v_{cut}$ for the majority of their history, but they can still have their supply of cold gas cut off by becoming a satellite of a larger galaxy, or temporarily cut off for a short period immediately after reionisation. In Figure~\ref{fig:mass_SMA}, these dwarfs are found in the two largest mass bins, with stellar masses of $10^6 - 10^9 $\Msun, equivalent to the M and L mass bins.
    \item `Turnover' dwarfs. These dwarfs tend to form around 50\% of their mass by $z \sim 2$, and are in the intermediate stellar mass range between $10^{4.5} - 10^6 $\Msun. As the critical halo mass for reionisation effects increases, these galaxies have haloes that, due to their assembly histories, fall below that mass at some point, even if initially they started out as larger haloes than are found in the M mass bin at reionisation. as shown in Figure~\ref{fig:RI_halogrowth}. They are not true relics, because they have higher stellar masses than galaxies whose growth is cut off at reionisation, but their halo growth falls below the $v_{cut}$-defined halo mass at an average time of $z\sim 2$, cutting off their ability to cool gas if they are still central galaxies. This group is very rare compared with the other two populations, and the exceptions to the rule of reionisation creating a bimodal mass distribution. In Figure~\ref{fig:mass_SMA} these are found in the second smallest mass bin, labelled S. In section~\ref{sec:types}, we also show that this mass group is an outlier in that it contains a population of mostly orphan galaxies.
    %\WH{Robustness check request (mass-bin drift): several environment trends (especially in the S bin) appear sensitive to the fact that the final $M_\star$ distribution differs between environments within a nominal bin (e.g. node objects drifting to lower $M_\star$ and looking more relic-like). Can we add a brief robustness statement: either (i) show that trends persist when matching the $z=0$ $M_\star$ distributions across environments (e.g. reweighting/subsampling), or (ii) explicitly caveat that part of the effect may be driven by within-bin mass differences? Even a short note pointing to a test (if already done) would pre-empt a referee selection-effect critique.} \Mac{I have an explicit caveat for this in the sections about environment trends, specifically looking at the S mass bin. It's true that the S mass bin specifically is super sensitive to slight changes in distribution, because there are so few central galaxies in it, and they tend to straddle the ``critical mass'' I have been talking about in terms of stellar mass. So, in Figure 10's S mass bin for centrals, you see the node galaxies don't have as many as the others, and the 50th percentile has pinged lower. You can see from the standard deviation that they don't have significantly lower masses than the other galaxies in that sense though. I chose to just discount this particular 50th percentile line from any overall trend calculation, because it was clear the final masses of the 50th percentile line were significantly lower.}

\end{enumerate} 

In Figure~\ref{fig:mass_SMA}, we see the XS mass bin, the ``Relic Dwarfs'', have assembled 50\% of their mass by $z = 6.6$, roughly 2.1 Gyrs before the S mass bin at $z = 2.3$, the ``Turnover Dwarfs''. The two largest mass bins, M and L, represent the larger mass dwarfs, which do not contain any reionisation relics and which continue to form stars over a longer period of time  (see also Figure~\ref{fig:RI_halogrowth} ). These mass bins assemble 50\% of their mass by $z = 0.52$ for L, and $z = 0.61$ for M. This is only a difference of 0.5 Gyrs, but this difference is constant across the average assembly histories of the two mass groups. The largest galaxy mass bin in the dwarf regime, L, therefore assembles 50\% of its mass, on average, roughly 7.7 Gyrs later than the smallest galaxy mass bin in the dwarf regime, XS.

Figure~\ref{fig:RI_halogrowth}, shows the mass assembly histories of the dark matter haloes hosting these galaxies. The black dashed line denotes evolution of the critical halo mass threshold above which gas cooling can take place (through atomic processes), starting at the atomic cooling limit and then jumping to the higher limit defined by reionisation at $z=6$. To construct this curve, we have assumed that only haloes with circular velocities above $\approx 17\,{\mathrm{kms}}^{-1}$ are able to cool gas (corresponding roughly the atomic cooling limit). After reionisation at $z=6$, this threshold is raised to $30\,{\mathrm{kms}}^{-1}$, corresponding to our fiducial choice of $v_{cut}$. The solid lines show the 50th percentile halo mass assembly for central galaxies in our different stellar mass bins seen in Figure~\ref{fig:mass_SMA}. Reionisation is imposed at $z=6$ in our fiducial model, cutting off star formation effectively in the XS mass bin ($M_{\star} = 10^{3} - 10^{4.5} $\Msun) due to them being unable to cool any new gas to form stars with. On average, these relic galaxies effectively have only 0.28 Gyrs in which to cool gas for star formation. Galaxies in the intermediate mass bins, S and M, experience a strong dip in their star formation at the beginning of reionisation, but recover and form more stars before $z=0$. In the S mass bin ($M_{\star} = 10^{4.5} - 10^{6} $\Msun),  galaxies have two distinct periods during which to cool gas for star formation, totalling 1.1 Gyrs. The next-largest mass bin, M ($M_{\star} = 10^{6} - 10^{7.5} $\Msun), galaxies, on average, cool gas over nearly 11 Gyrs. This is a significant difference in the duration of star formation in the lifetimes of these objects, despite the halo masses of both the S and M galaxies being somewhat similar.

In the region of the two medium mass bins, the stellar-halo mass relation is extremely steep (Figure~\ref{fig:mstars_mhalo_fiducial}), meaning that central galaxies of very different stellar masses are hosted inside dark matter haloes of the same mass. This indicates that while the haloes may end up at a similar mass at final day, the rate at which they grow just after reionisation is what results in very different galaxy populations by $z=0$. In the M mass bin, the halo continues to grow post-reionisation, whereas in the S mass bin, the growth hits a plateau, preventing the galaxy within from acquiring more cool gas. This is the explanation for the steepness of the stellar-halo mass relation in this region: very small changes in halo mass at any time after reionisation can either result in periods of extended or very brief star formation.

In the largest mass bin, L, where $M_{\star} = 10^{7.5} - 10^{9} $\Msun, the halo mass is never below the critical mass scale for gas cooling before or after reionisation. Their host galaxies therefore experience largely uninterrupted gas cooling, resulting in continuous star formation throughout their lifetimes.

In these figures, we show only central galaxies' stellar mass assemblies within each mass bin. As we will show in the next subsection, the type of each galaxy has a profound impact on the stellar mass assembly of these mass bins.

\subsubsection{Galaxy Type: Centrals, Satellites and Orphans}
\label{sec:types}
\begin{figure}
	\includegraphics[width=\columnwidth]{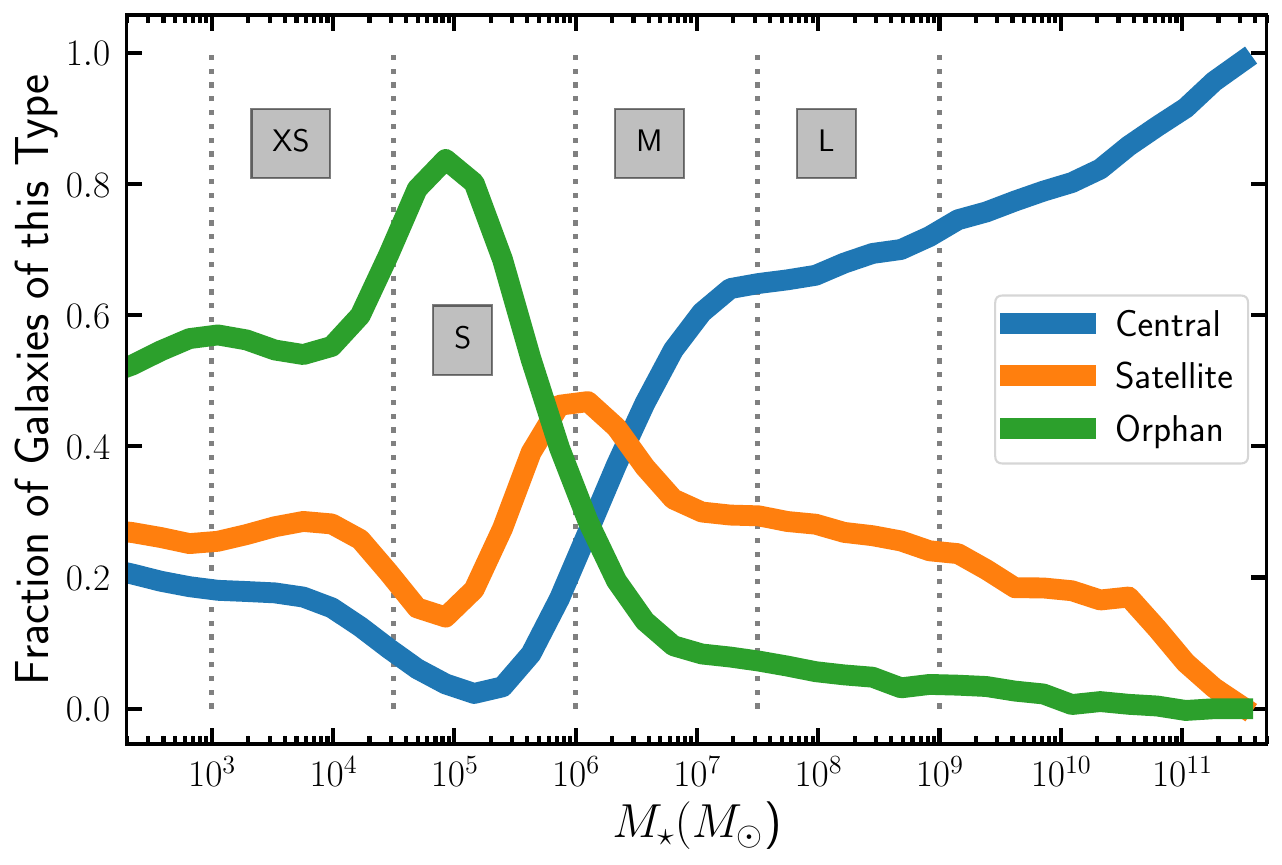}
    \caption{This figure shows the fraction of galaxies, at a given mass, of each type. Galaxies where $M_{\star} < 10^5 $\Msun{} are mostly orphan galaxies, galaxies where $M_{\star} \sim 10^6 $\Msun{} are a mix, with satellites making up the majority, and galaxies where $M_{\star} > 10^{6.5} $\Msun{} are majority centrals, with very few orphans. The peak in orphan fraction coincides with a dip in the stellar mass function at around $M_{\star} = 10^5 $\Msun{} (see Figure~\ref{fig:SMF_fiducial}). The most massive galaxies, with $M_* \geq 10^{10} $\Msun, are overwhelmingly central galaxies. The different types found in each mass bin have a profound effect on the stellar mass assembly of the galaxies (see Figure \ref{fig:type_sma}).}
    \label{fig:galtype_frac}
\end{figure}

%\WH{Physics wording check: the text currently states satellites are `quenched soon after entering' their host halo, but later we discuss evidence for a 2--4~Gyr delay between infall and quenching (Wetzel+2013; Rintoul+2025) and mention Hou+2018 improvements. Can we align the Results wording with the Discussion by making clear what `instant ram-pressure stripping' means in the fiducial GALFORM implementation (hot-halo removal vs immediate SF shutdown), and what timescale for SFR decline is implied in the model? This would avoid giving the impression that GALFORM predicts immediate quenching.}
%\Mac{To be clear: the fiducial GALFORM model used in this paper includes instant ram-pressure stripping, but not instant star formation shutdown, because the cold gas currently in the galaxy can still be used up. I am not 100\% sure what the average timescale is for the cessation of SF once the galaxy becomes a satellite, but I can find out a rough answer. The problem is that we don't know the exact time when a galaxy became a satellite in GALFORM, just the last snapshot at which it was identified as a central galaxy. So, this calculation will have some complicated systematic errors on it.}
The next step in measuring the effect of environment on dwarf galaxies is to examine the effect of their placement in their dark matter haloes. If they are the largest galaxy in their dark matter halo, they are a central galaxy, and therefore able to access all the gas in the local environment surrounding that halo. If they are not the largest galaxy in their dark matter halo, they are a satellite galaxy. These do not have access to gas from the intergalactic medium, and are quenched soon after entering the larger dark matter halo.

Orphans exist when a galaxy once had a large enough dark matter halo to be resolved in COCO, but due to tidal disruption, its parent halo drops below the resolution limit of the simulation. When this happens, GALFORM tracks the most bound particle and labels the galaxy as an orphan. This represents what is likely a real group of galaxies that are disrupted by other galaxies, and may survive as very small, closely-orbiting satellites \citep{santos-santos2024}.

In Figure~\ref{fig:galtype_frac}, we can see what types of galaxies are most common at different masses. For reionisation relics and turnover dwarfs (the XS and S mass bins), orphan galaxies dominate. For larger dwarfs, centrals dominate. The peak in the percentage of orphan galaxies occurs at $M_{\star} = 10^{5} $\Msun, where about 80\% of the galaxies are orphans. At $M_{\star} = 10^{6} $\Msun, there is a peak in the percentage of satellite galaxies, at around 50\%. Centrals make up more than 50\% of the galaxies from $M_{\star} = 10^{6.7} $\Msun upwards, with centrals accounting for 80\% of galaxies over $M_{\star} = 10^{10} $\Msun. The stellar mass at which the orphan population peaks is set by the choice of $v_{cut}$ and $z_{cut}$, as it is driven by a lack of non-orphan galaxies at this mass, also visible as a trough in the stellar mass function (see Figure~\ref{fig:SMF_fiducial}, Figure~\ref{fig:nexusSMF}, and exploration of this feature in the stellar mass function in \citet{bose2018}). It represents the transition between the mass scale where the effect of reionisation dominates the growth of the galaxies, and the masses at which galaxies are no-longer reionisation relics. The mass scale of $10^5$ \Msun{} corresponds to the characteristic stellar mass, below which, haloes are affected by reionisation. The galaxies in this regime are preferentially orphans.  We find that at $z=6$, they were large enough to not become true reionisation relics, but their abnormally early infall into a host halo stopped their growth, preventing them from forming enough stars to fall into the higher mass bins (see Figure~\ref{fig:infall}).

\begin{figure*}
	\includegraphics[width=\textwidth]{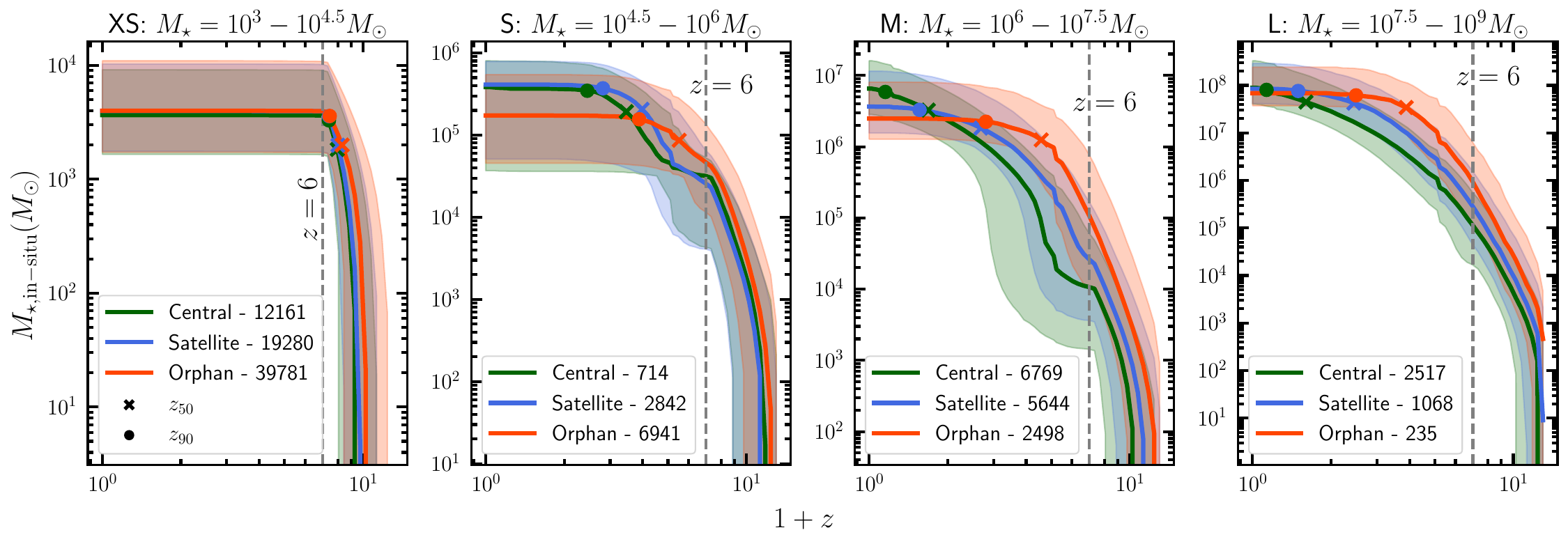}
    \caption{The in-situ stellar mass growth for galaxies in COCO/GALFORM, divided into $4$ mass bins, and then categorised by their galaxy ``type'' (central galaxies, satellite galaxies, or orphan galaxies). The solid line shows the 50th percentile of the mass bin, while the shaded regions show the 16th to 84th percentiles. The crosses along the solid lines denote where the galaxy reaches 50\% assembly, while the dots represent 90\% assembly. The dashed vertical line at $z=6$ shows the timing of reionisation. Central galaxies keep assembling stars for longer than satellite or orphan galaxies, regardless of the mass range. Orphans finish their assembly first, and even in the highest mass bin, orphans have on average finished their assembly before $z \sim 2$. In the smallest mass bin, we see the smallest differences between the different galaxy types because the average satellite infall time is generally later than $z = 6$, when reionisation causes most or all galaxies within this mass bin to quench. So, by $z = 6$, only a small percentage of final-day satellites will have fallen into their host haloes, making this the mass bin with the least difference between galaxy types.}
    \label{fig:type_sma}
\end{figure*}

Figure~\ref{fig:type_sma} shows the stellar mass assembly of different galaxy types within the mass bins discussed in Section~\ref{sec:mass_ass}. The leftmost panel of Figure~\ref{fig:type_sma} is the XS stellar mass bin, from $10^3 - 10^{4.5} $\Msun. These galaxies are almost exclusively reionisation relics, galaxies which have small dark matter haloes which are below the circular velocity at $z=6$ for them to cool additional cold gas into their haloes. In the smallest mass bin, we see the smallest differences between the assembly of different galaxy types, with central galaxies reaching 50\% assembly 0.05 Gyrs later than orphan galaxies. This is because the average satellite infall time is later than $z = 6$ (see Figure~\ref{fig:infall}), when reionisation causes most or all galaxies within this mass bin to quench. By $z = 6$, only a relatively small percentage of final-day satellites and orphans will have fallen into their host haloes. Reionisation at $z = 6$ prevents further star formation, and therefore in-situ stellar mass assembly, before the types of galaxies found at $z = 0$ to be satellites or orphans have had a chance to become satellites or orphans, so the assembly of the different galaxy types is the most similar in this mass bin.

%\WH{Robustness check request: since the orphan population depends on the subhalo disruption criterion and effective resolution, can we confirm that the main qualitative conclusions (especially the environment trends) do not hinge on orphans? If feasible, it would be useful to state explicitly whether key trends persist when excluding orphans, or to add a brief caveat if not tested.}
%\Mac{To be clear, Orphans are not included in our environment classifications because they are, by definition, not recognised as haloes at $z=0$. So, the results as they are are without orphans being considered. We could potentially test adding them in, but I think this would take more time than I have left for this paper. So, I think I will add a clearer caveat that orphans are not tested when looking at LSS environments, or at Tilly's local environment measure.}
The middle-left panel is the S mass bin, made up of majority orphan galaxies, with a smaller number of satellites and a much smaller number of central galaxies. Here, there is a dramatic difference when compared with the XS mass bin, as the vast majority of the mass assembly of these galaxies happens after $z = 6$. We also see a larger difference in the evolutionary paths for these objects based on type, with orphans the most varied but also the fastest to finish their assembly, and a much larger difference between satellites and central galaxies when compared with the smallest mass bin. In this mass bin, central galaxies assemble 0.5 Gyrs later than satellites, and 1.4 Gyrs later than orphans. Orphans are also on average the smallest galaxies in this mass bin, with their final-day masses significantly smaller than central galaxies. The average final-day stellar mass of the orphan galaxies in this bin is less than half of the final day masses of the central galaxies. This is due to the fact that there is a peak in the proportion of orphan galaxies at around $10^5 $\Msun, as is shown in Figure~\ref{fig:galtype_frac}.

%\WH{Physics consistency check: please confirm the intended interpretation of `orphan' evolution in this analysis. In Sec.~2.2 we state that once a galaxy becomes an orphan it cannot evolve further within GALFORM (except for merging), but in Sec.~3.1.2 we describe orphans as `continuing to form stars' and being less impacted by reionisation. Are we referring to (i) their growth \emph{prior} to orphanisation (e.g. earlier infall, higher pre-infall mass, hence weaker reionisation suppression), or (ii) continued star formation \emph{after} becoming an orphan? If (ii), we should clarify how GALFORM implements SF for orphans to avoid an apparent contradiction.}
%\Mac{Definitely referring to pre-orphanisation here. Once they become orphans, as far as I know, their evolution totally stops and GALFORM just tracks the most-bound particle. I am saying here that prior to becoming orphans, they had higher masses than "true" reionisation relics, which then they were frozen at those masses soon after.}
In the two middle mass bins, S and M, we can see an interesting pattern emerge in the assembly of the different types of objects. We can see that orphan galaxies are not only building up 50\% of their masses earlier than other galaxies, but that orphan galaxies are actually more massive than satellite and central galaxies at $z = 6$. This means that orphans are less impacted by reionisation, and continue to form stars, whereas the satellite and central galaxies undergo a post-reionisation ``dip'', before recovering and surpassing the orphans in mass by $z=0$. This is because we have selected for galaxies which end their assembly with the same masses, but orphans infall into their host haloes (on average) earlier than satellites or central galaxies. Therefore, to end up with the same mass, the orphans must have been larger in the distant past to make up for their earlier star formation cut-off due to infall.

The two largest mass bins, M and L, are similar in assembly, with almost all the mass assembly occurring after $z=4$. In the M mass bin, central galaxies form 50\% of their stellar mass 3.6 Gyrs later than satellites, and 5.7 Gyrs later than orphans. This is similar to the difference in assembly timescales between centrals, satellites, and orphans in the largest mass bin, L,  where there is also a decreased percentage of satellite and orphan galaxies. The orphans and satellite galaxies within this mass bin have, on average, later infall times in order to build up enough stellar mass to qualify for this mass bin (this is shown later in Figure~\ref{fig:infall}). This later infall time is responsible for the slight decrease in the magnitude of the mass assembly trends by galaxy type within this mass bin.% when compared with the second-largest mass bin, especially because of the lack of the effect of reionisation in this mass bin, as mentioned when discussing the M and S mass bins.}

As discussed in Section~\ref{sec:galform}, the GALFORM treatment of satellite galaxies prevents gas from cooling into their dark matter haloes, and as a result, satellites assemble their mass earlier than central galaxies of the same final day mass. The treatment of orphan galaxies is similar, in that they are prevented from evolving after moving below the halo resolution limit. Additionally, galaxies with a circular velocity $v_{circ} < 30{}$km{}~s$^{-1}$ are prevented from cooling gas into their haloes after $z=6$, mimicking the effect of the UV background and reionisation.

Overall, we see significant trends, in all mass bins, of orphan galaxies halting their assembly first, then satellites, and central galaxies assembling last. There is also a sweet spot in the mass range for seeing these trends, caused by the infall time of the satellites and orphans for different mass bins, as well as the selection effects created by reionisation's impact on galaxies at lower masses. For the lowest mass galaxies, the trends by galaxy type are much smaller due to reionisation cutting off star formation in the mass bin well before a significant number of the final-day orphans and satellites have actually fallen into their host haloes. On the other hand, in the largest mass bins, the typical infall times for satellites and orphans is delayed, almost by definition. This, in turn, shortens the duration over which environmental effects are able to significantly impact these galaxies by $z=0$. The infall time of the satellites will also be important later, when we examine the trends with respect to the large-scale structure. %See Figure~\ref{fig:infall} and Section~\ref{sec:differencescauses} for the explanation for this.

\subsubsection{Large Scale Structure}
\begin{figure*}
	\includegraphics[width=16cm]{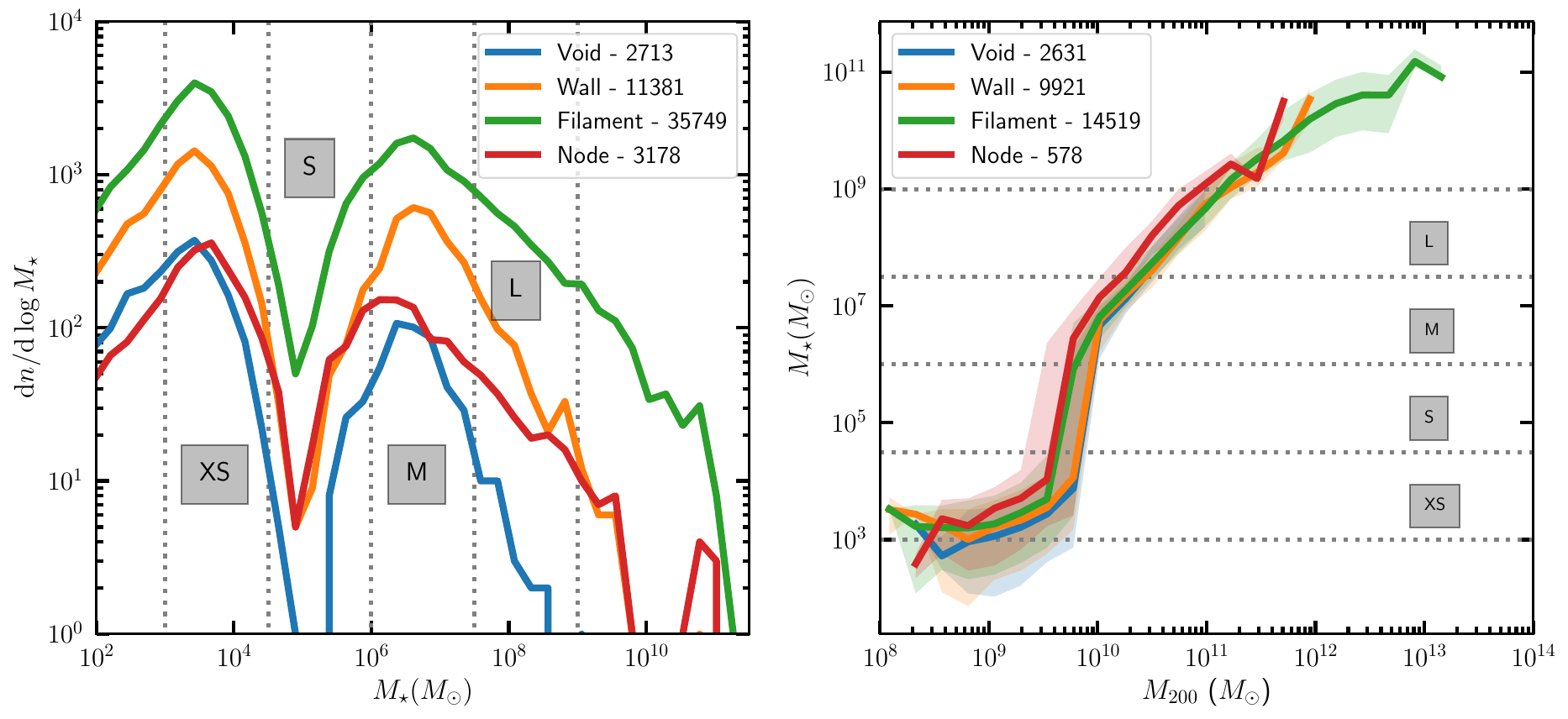}
    \caption{The differential stellar mass functions of all galaxies for different regions in NEXUS+ (left panel), and the stellar-halo mass relations for central galaxies of the different regions (right panel). The galaxy stellar mass function (left panel) is bimodal for all regions in NEXUS+. The total number of galaxies in each region is noted in the legend, with filaments containing more galaxies than the other regions combined. Void regions do not contain any galaxies above $M_{\star} > 10^{9}$\Msun. For the stellar-to-halo-mass relation (right hand panel), the shaded regions show the 16th to the 84th percentiles for the relation. The lines for each region show that in the denser regions, galaxies have higher stellar masses for any given halo mass, and the opposite is true for the less-dense regions. However, the overall shape of the relation is unchanged regardless of environment.
}
    \label{fig:nexusSMF}
\end{figure*}

\begin{figure*}
	\includegraphics[width=\textwidth]{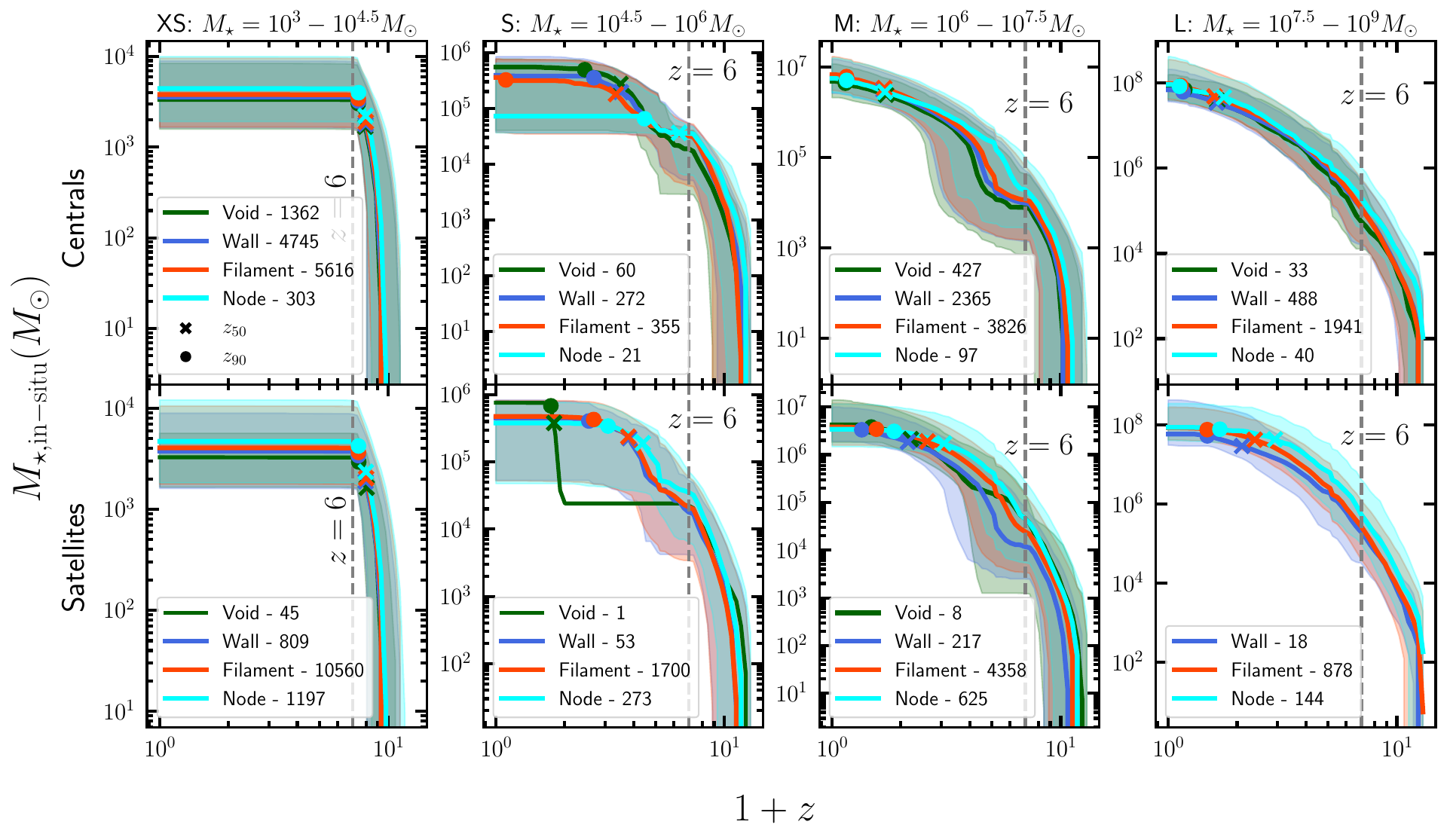}
    \caption{In-situ stellar mass assembly for galaxies divided into mass bins (columns) and into types (rows), and further divided into regions of large scale structure within the panels. The x-axis is the redshift, while the y-axis shows the in-situ stellar mass at each redshift. Solid lines represent the 50th percentile of each sample, and the shaded regions represent the 16th to 84th percentiles within that sample. Over all the mass ranges, node galaxies finish assembly earlier than wall and void galaxies. We see very small trends between environments in the central galaxies, with an average $z_{50}$ assembly time 0.69 Gyrs different between the void and node regions, and when removing the outlier nodes group in the S mass bin, whose final stellar masses are on average below $10^{5}$\Msun, this decreases to 0.08 Gyrs. The trends in central galaxy assembly are biggest in the epoch of reionisation, because the haloes that node galaxies form in are larger than the other galaxy haloes at this time. These trends disappear after reionisation, because the haloes in node regions do not grow at the same rate as haloes in sparser regions at later times, allowing the stellar mass assembly of the other regions to ``catch up'', post-reionisation. The trends in the satellites for stellar mass assembly are far greater, with 1.5 Gyrs being the average difference between $z_{50}$ in wall and node satellites. The stellar mass assembly differences quoted here only take into account the three most massive mass bins, due to the lack of discernible trends in the smallest mass bin, for reasons already commented on in Figure~\ref{fig:mass_SMA}.}
    \label{fig:nexusSMA}
\end{figure*}

Having examined the dependence of the stellar mass assemblies on the galaxy type (centrals vs. satellites vs. orphans), we next investigate the connection with the large-scale structure in which these galaxies are embedded. In this paper, we consider $4$ different regions of the cosmic web: Nodes (the most dense), Filaments, Walls and Voids (the least dense). For a visual representation of these environments, see Figure~\ref{fig:coco_pic}. For more details on how these environments are identified,  see Section~\ref{sec:Nexus}.

%WH{Sec.~3.1.3: add a short clarification that cosmic-web trends partly reflect differences in central/satellite/orphan mix across environments, and that later we explicitly isolate satellites to interpret assembly-time trends via infall-time distributions (Fig.~13). This avoids readers thinking we are comparing like-with-like in every panel by default.} \Mac{Yeah I agree this is helpful.}
% below paragraph added by Wojciech
We note that trends with cosmic-web environment can reflect both genuine differences in assembly histories at fixed mass and differences in the mix of galaxy types (centrals, satellites, and orphans) across environments. Where relevant below, we therefore separate results by galaxy type, and in particular we later focus on satellites when interpreting assembly-time trends via the distribution of satellite infall times (Figure~\ref{fig:infall}).

%The goal of this project was to identify whether our combination of GALFORM and COCO can produce differences in the stellar mass assembly of dwarf galaxies, despite GALFORM lacking any environment-based dependencies for galaxy formation. Stellar mass is impacted by the large scale structure environment, as indicated in Figure~\ref{fig:nexusSMF}. 
The left panel of Figure~\ref{fig:nexusSMF}, shows the stellar mass function of galaxies split by regions of the cosmic web. There are opposing effects due to the mass scale that contribute to the percentages of the galaxies in each region that are ``relics''. For denser environments, there is more opportunity for halo growth through mergers with other galaxies, and smooth accretion of the denser dark matter environment. However, not all galaxies will be the beneficiaries of this increased halo mass, as some will have their star formation suppressed due to becoming satellite or orphan galaxies. %\WH{Clarification request: please state explicitly the operational definition of a `reionisation relic' used throughout the paper (e.g. a cut on final stellar mass, a criterion on post-reionisation star formation, a flag internal to GALFORM, or another definition). Since several key statements hinge on relic fractions by environment/type, the definition should be spelled out once, clearly, and then used consistently.} \Mac{The definition used here is galaxies below $10^5$ in stellar mass, but in the rest of the paper it's a more general "on the left side of the bimodality" rule, which has a dip at $10^5$ anyway. Corresponds to galaxies who form $>70\%$ of their mass prior to RI \citet{bose2018}.}

%\WH{Sec.~3.1.3 (Fig.~9 discussion): after stating the relic fractions by environment, explicitly note that the central/satellite composition differs strongly between environments (e.g. void relics are almost exclusively centrals), which is itself part of the physical picture and also motivates later focusing on satellite-only trends when discussing infall-driven effects.} \Mac{Yep, makes sense.}
In the left panel of Figure~\ref{fig:nexusSMF}, we can see that in node and filament regions, 64\% and 63\% of galaxies have $M_{\star} < 10^5$\Msun, so the two environments are similar in this respect. There are much higher percentages of relic galaxies in the void and wall regions, 81\% and 71\% respectively. This indicates that the higher relic fractions in voids and walls are closely tied to a different present-day galaxy-type mix in these environments. Indeed, non-orphan relic galaxies in walls are 86\% centrals and in voids are 100\% centrals, whereas in nodes only 20\% are centrals, with 80\% having become satellites by $z=0$. This supports interpreting the strongest assembly-time differences through satellite-specific processes and their infall histories, while in the most underdense environments a larger fraction of systems remain isolated and experience limited late-time growth.
%\sout{This indicates that in those regions, isolation is driving a lack of growth for small haloes, which persists up until the present day, resulting in a higher fraction of galaxies in those regions being reionisation relics. Indeed, non-orphan relic galaxies in walls are 86\% central galaxies, and in voids are 100\%, whereas in nodes, only 20\% are central galaxies.}
% above paragraph edited and re-written by Wojciech

In the right-hand panel of Figure~\ref{fig:nexusSMF}, we can see the stellar-to-halo-mass relations of central galaxies in the different regions. We find that central galaxies in nodes have systematically higher stellar masses for a given halo mass. The median node galaxy in the XS stellar mass bin has a halo mass that is $\sim45$\% lower in mass than the corresponding void galaxy. This trend is roughly consistent across the mass bins.%, with the S mass bin having a 50.1\% difference, the M mass bin having a 30.7\% difference and the L mass bin having a 12.2\% difference. 
The differences in the stellar mass content of haloes in different environments is related to when the haloes grew. Haloes in nodes, on average, form earlier than those in other regions of the cosmic web, and were larger than them at early times, and so they built up more stellar mass while the other haloes were being impacted by reionisation, and before the other haloes had begun to form stars. They were more consistently above the threshold for the atomic cooling limit and reionisation, because their growth largely happened earlier, and slowed down later, making them the same masses as haloes in less-dense regions which formed later, but are growing quickly.

%This is in agreement with results found in \citet{zehavi2018} and \citet{celesteartale2018}, and are the result of mass assembly bias. However, these haloes couldn't continue to build up halo mass at the same rate forever, because they would then qualify for larger stellar mass bins, and so we have selected for haloes which grew less at later times, resulting in galaxies with a higher abundance of stellar mass to their halo masses for node galaxies.}

%\WH{Clarification request: please specify how $z_{50}$ is defined and measured in this work. Is it the lookback time/redshift when the galaxy assembled 50\% of its \emph{final} $z=0$ stellar mass, or 50\% of its \emph{peak} stellar mass (if stripping occurs), and how are orphans handled in this computation? This matters for interpreting environment trends, especially if tidal stripping can reduce $M_\star$ after infall.} \Mac{$z_{50}$ is based on the final stellar mass AND the peak stellar mass. There is no stellar mass stripping encoded into GALFORM. For orphans, their final stellar mass is also their peak stellar mass. Isabel's most recent version of GALFORM includes stripping, but this version doesn't.}
In Figure~\ref{fig:nexusSMA}, we find there are clearly differences in the stellar mass assembly of galaxies according to their environments, specifically, that in the majority of panels, node galaxies finish their assembly earlier than wall and void galaxies. For the top row, showing central galaxies, the differences between different environments are reduced compared to the results for satellite galaxies in the lower row. For central galaxies, the time at which 50\% of the stars in the galaxy are assembled ($z_{50}$), is the most different between galaxies in the S mass bin ($M_{\star}=10^{4.5} - 10^{6} $\Msun) with a 1.6 Gyrs difference between the void and node galaxies, with nodes assembling first. This trend is not due to the effect of the galaxy environments alone though, and is largely driven by the fact that the average final stellar mass of galaxies in node environments in this mass bin is below $M_{\star} = 10^5$ \Msun, meaning they are relic galaxies, whereas the average final stellar masses of the galaxies in the other environment bins are significantly above $10^5$\Msun, making them not relic galaxies. The S mass bin, for central galaxies, is extremely sensitive to small differences in the stellar mass function of different regions due to the bimodal nature of the stellar mass function. When comparing the difference between the assembly times of the more similar final mass filament and void galaxies, the difference between their assembly in Gyrs becomes reversed: 0.19 Gyrs but with the lest dense environments finishing their 50th percentile assembly first, making the overall trend of central galaxy assembly being that galaxies in denser environments assemble 0.08 Gyrs earlier than galaxies in the least-dense environments. %\WH{Clarification request: please specify the operational definition of `infall time' used in Fig.~13 and in the Discussion (e.g. first time the galaxy becomes a satellite in the merger tree, first crossing of the host $R_{200}$, time of peak halo mass, or another criterion). Since the physical interpretation hinges on the quoted $\sim 5.2$~Gyr difference, we should ensure this definition is explicit and consistent across the text and caption.} \Mac{Infall time here is the last snapshot at which the galaxy was ID'd as a central in GALFORM. This comes from the merger trees, which come from an FOF algorithm. So, once the DM halo is no-longer ID'd in the FOF algorithm as it's own thing, it's then a satellite in the merger tree. GALFORM runs on this merger tree.}

Differences in the number of galaxies of certain types within different mass bins in Figure~\ref{fig:nexusSMA} are due to environmental effects on the haloes of the galaxies. For example, there are only 54 satellites in void regions, compared to hundreds in nodes, despite the fact that they have a comparable total number of galaxies. This is because, for an object to become a satellite, it must be near something larger than itself. In a region with many fewer interactions between galaxies, galaxies are less likely to have massive neighbours. This also means that there is an increased likelihood of small galaxies being satellites, because if they have an interaction, it is more likely to be with a more massive galaxy.  Considering this, the different proportions of galaxy types in each NEXUS+ environment is not concerning.

Satellite galaxies have larger differences than central galaxies between the types of environment in a given mass bin, with node galaxies assembling significantly earlier than wall, filament and void galaxies, and the other environments also following the trend of higher density leading to earlier assembly. The average $z_{50}$ assembly difference between the node and wall regions for satellites is 3.0 Gyrs, but this is reduced to 1.5 Gyrs when discounting the S mass bin void satellite which is the only member of its group, and so unlikely to be representative. Given that GALFORM has no density/spatial-dependence in how reionisation is implemented these differences in satellite assembly showing stronger environmental trends has to do with the time at which satellites first infall into a host halo.

The mass bin with the largest difference in $z_{50}$ across environments is the M mass bin, where $M_{\star} = 10^{6} - 10^{7.5} $\Msun. In this mass bin, the wall satellite galaxies assembled 50\% of their mass 2.2 Gyrs later than the node satellite galaxies. In a sense, this mass bin acts as a ``sweet-spot'' for gauging environmental differences: at the more massive end, the selection by final day stellar mass preferentially selects objects that fall into their hosts much later, whereas in the lowest mass bins, reionisation quenches star formation in dwarfs long before any environmental effects become important. Both effects conspire to minimise the differences between centrals and satellites in different regions of the cosmic web.

%The reason for the mass dependency in the trends in satellites is due to, at the more massive end, the selection bias leading to their infall time being later, and at the smallest end, their star formation being too disrupted by reionisation, even before they fall into their host galaxy halo. This idea is explained in more detail in Section~\ref{sec:differencescauses}, with Figure~\ref{fig:infall}.}

The next step for our investigation is to ascertain whether these trends can also be replicated when using a different measure of environment, such as the local density of the dark matter haloes. This will be the point of investigation in the following subsection.

\subsubsection{Local Density of DM Haloes}
\label{sec:localdm}
\begin{figure*}
	\includegraphics[width=\textwidth]{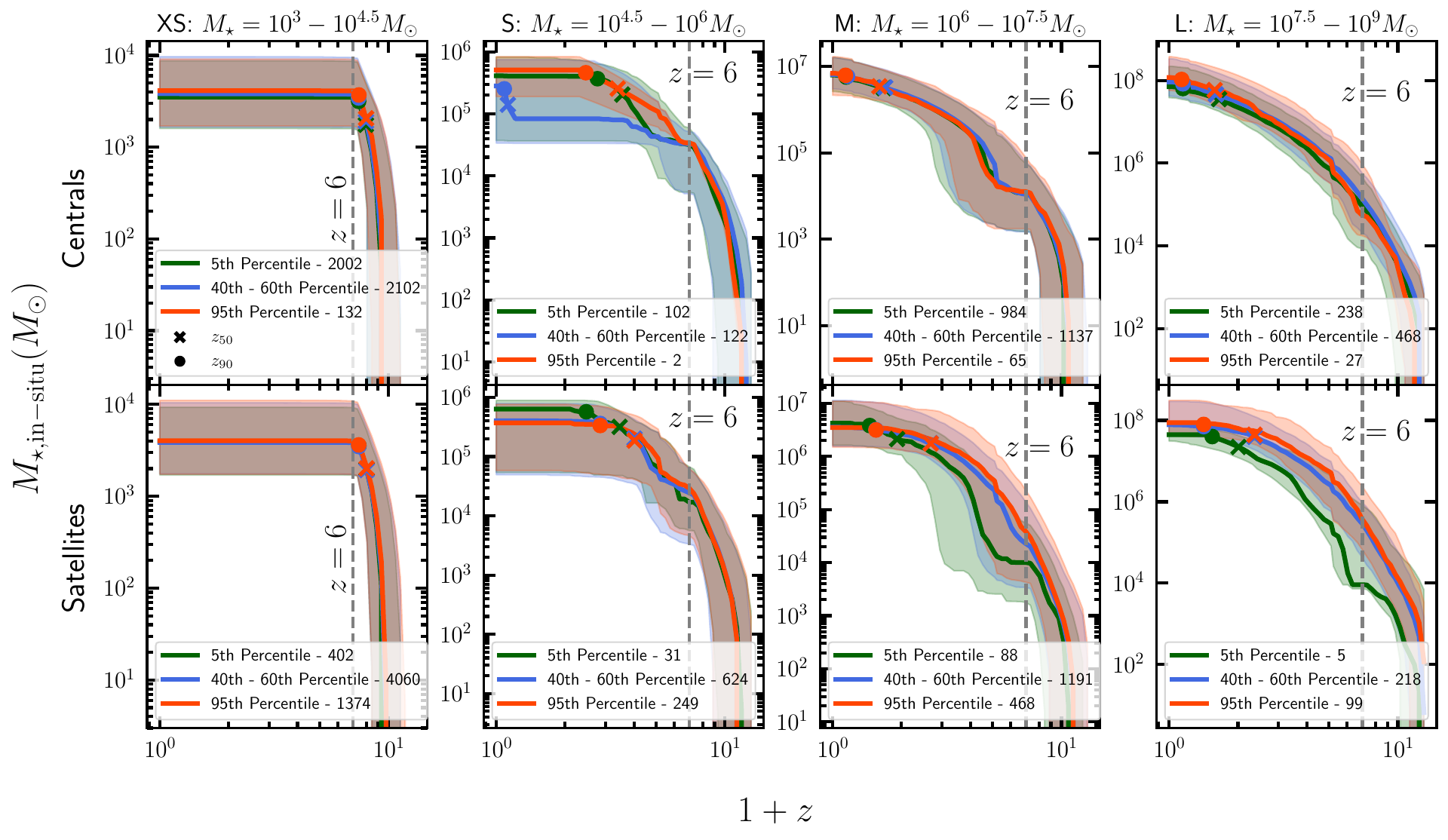}
    \caption{Stellar mass assembly for different galaxies, categorised by mass and by the local density environment measure (for more information on this, see Section~\ref{sec:localdm}). The y-axis shows the in-situ stellar mass, and the x-axis is z+1, represented on a log scale. The solid lines show the 50th percentile of each sample, and the shaded regions show the 16th to the 84th percentiles of the sample. Again, the trend is found that the central galaxies appear to have very small trends, with 5th percentile galaxies reaching $z_{50}$ 0.37 Gyrs before the 95th percentile galaxies. The satellite galaxies have large differences in their assembly histories depending on the density of their environment, with the 95th percentile galaxies reaching $z_{50}$ 1.4 Gyrs before the 5th percentile galaxies. The smallest mass bin shows negligible trends with environment due to the reasons discussed in Section~\ref{sec:differencescauses}, with Figure~\ref{fig:infall}, which is that their star formation is cut off by reionisation before satellites infall into host haloes.}
    \label{fig:tilly_sma}
\end{figure*}

We define the local density of dark matter haloes by counting the number of subhaloes located in a shell surrounding each GALFORM galaxy, with an inner radius of 200 kpc and an outer radius of 2 Mpc. This definition quantifies the abundance of subhaloes in the immediate vicinity of each galaxy, whilst excluding any satellites that may be associated with it. We then define objects located in different environments by splitting them in percentiles of this statistic.
%This method measures the local density of dark matter haloes in the area of a given galaxy. The count associated with each galaxy is the number of subhaloes within $2$Mpc but excluding those within $0.2$Mpc (ie. satellites). This allows us to measure more directly the local density surrounding a galaxy, rather than relying on the overall density of the local area.

In Figure~\ref{fig:tilly_sma}, as expected, we reproduce the trends seen in the high and low density regions in NEXUS+. Satellites display stronger trends than central galaxies, and central galaxies show very small trends. The average $z_{50}$ difference between satellites in the highest density regions (95th percentile in density) and the lowest density regions (5th percentile in density) is 1.4 Gyrs, with satellites in the highest density regions forming first. Again, the environmental dependence in the smallest mass bin is negligible due to reasons discussed in Section~\ref{sec:Nexus} and Section~\ref{sec:differencescauses}. Comparing with the satellite $z_{50}$ difference in the NEXUS+ environments, which was 1.5 Gyrs, the difference in the local environment method is slightly smaller, but not significantly so.

In the three most massive bins, central galaxies show an average $z_{50}$ difference of 0.37 Gyrs, with galaxies in the least-dense environments finishing their formation later, a reversal in the trends we see in satellites, which appears to be driven by higher-density environment galaxies having slightly larger final-day stellar masses. In the lowest mass bin, these differences are vanishingly small because all of these galaxies are reionisation relics. There are too few central galaxies in the S mass bin to determine any trends.

Although there are small differences in quantitative detail between the two environment definitions, the overall qualitative trends are consistent between the two. This is to be expected, as even the more sophisticated environmental classification method of NEXUS+ is ultimately a reflection of the underlying dark matter density field.

\subsubsection{Causes of Environmental Differences}
\label{sec:differencescauses}
\begin{figure*}
	\includegraphics[width=\textwidth]{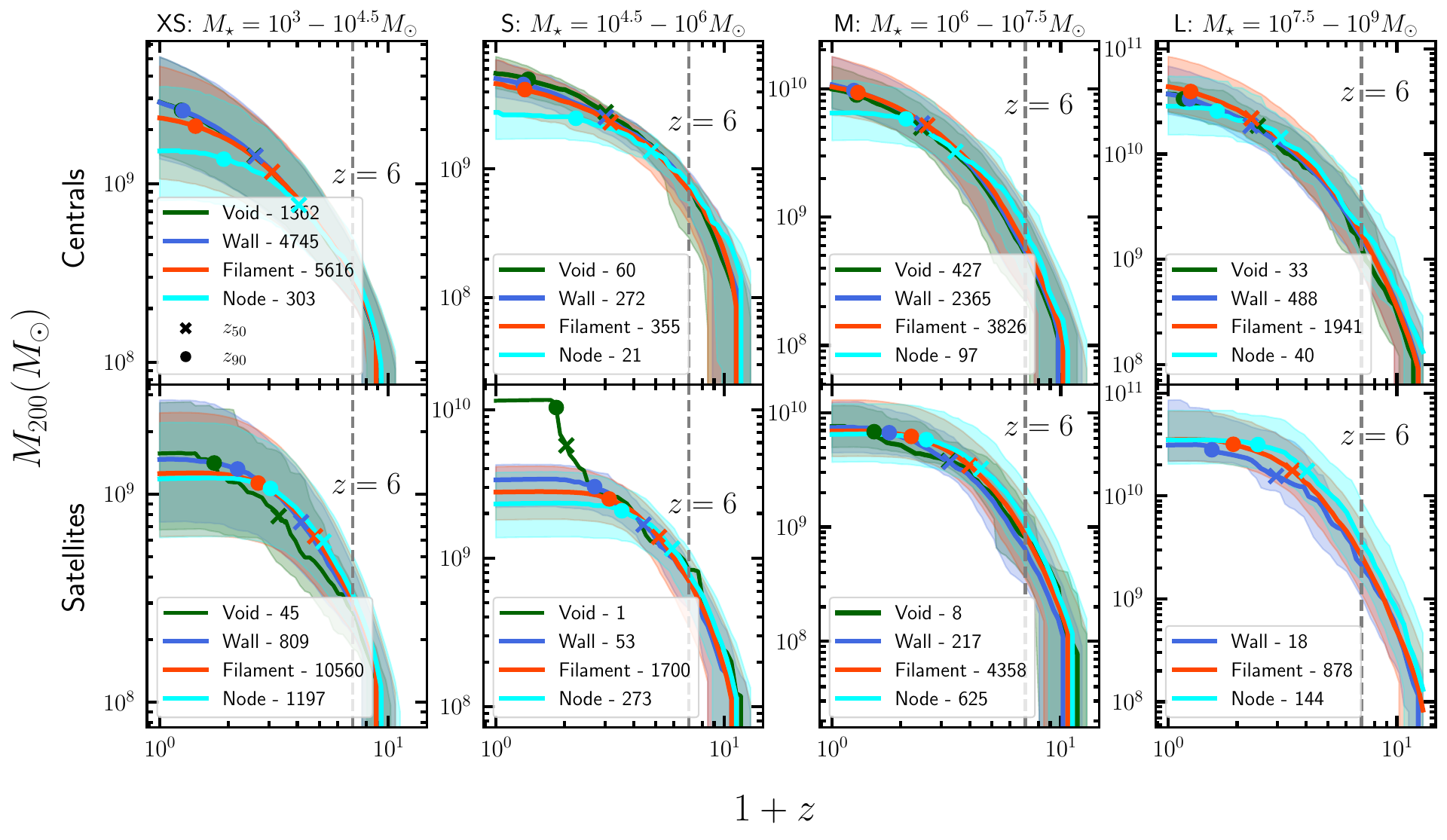}
    \caption{The halo mass assembly of galaxies divided into their NEXUS+ environments and mass bins. The solid lines represent the 50th percentile of each group of galaxies, and the shaded regions represent the 16th to 84th percentiles. Central galaxies show stronger environmental trends (1.5 Gyrs $z_{50}$ difference between nodes and voids) than those we see in the stellar mass assembly of central galaxies (0.08 Gyrs $z_{50}$ difference between nodes and voids). Satellite galaxies show much smaller trends in halo mass assembly (1 Gyrs $z_{50}$) than in stellar mass assembly (1.5 Gyrs $z_{50}$). This indicates that the environmental differences in satellite galaxy assembly are not explained by the trends in halo mass assembly between denser and less-dense areas. In fact, it indicates that in spite of the environmental trends in halo mass assembly, central galaxies in different environments do not assemble at appreciably different times. For satellites, the halo mass is defined the mass of the host halo {\it before} it became a satellite, which is then frozen once it falls in to a larger host.}
    \label{fig:nexus_halo_SMA}
\end{figure*}

Figures \ref{fig:nexusSMA} and \ref{fig:tilly_sma} both show that there is a large environmental dependency of the stellar mass assemblies of satellite galaxies, but a much smaller environmental dependency of central galaxies. Because GALFORM only considers the merger trees of a given final galaxy for its semi-analytic calculations, this means that the answer to what is causing the differences in environment, and the differences between satellite and central galaxies, must relate to the haloes and merger trees of the galaxies. Considering this, we examined the halo mass assemblies of galaxies in different regions, and the differences between central and satellite galaxies in Figure~\ref{fig:nexus_halo_SMA}, again divided into our 4 mass bins. The halo mass assembly shows small differences between the environmental groups. Haloes in denser dark matter regions (e.g. nodes) form first, as expected from hierarchical formation, by a significant amount for central galaxies (1.5 Gyrs difference in $z_{50}$ between nodes and voids), and a smaller amount for satellite galaxies (1 Gyrs difference in $z_{50}$ between nodes and voids). Node galaxy haloes are also growing more rapidly earlier on, and their growth also plateaus earlier. This is because while node galaxy haloes seem to assemble slightly earlier than other region's haloes, they are also significantly smaller for galaxies in the same stellar mass bins. The node haloes form earlier, allowing the galaxies to build up more stellar mass in smaller haloes, and allowing them to continue forming stars during the initial post-reionisation pause in star formation that lower-mass haloes experience. They then plateau because if they were to continue to build up mass and become the same masses as the haloes in other regions, they would have much higher stellar masses, exempting them from our stellar mass bins.

\begin{figure}
	\includegraphics[width=\columnwidth]{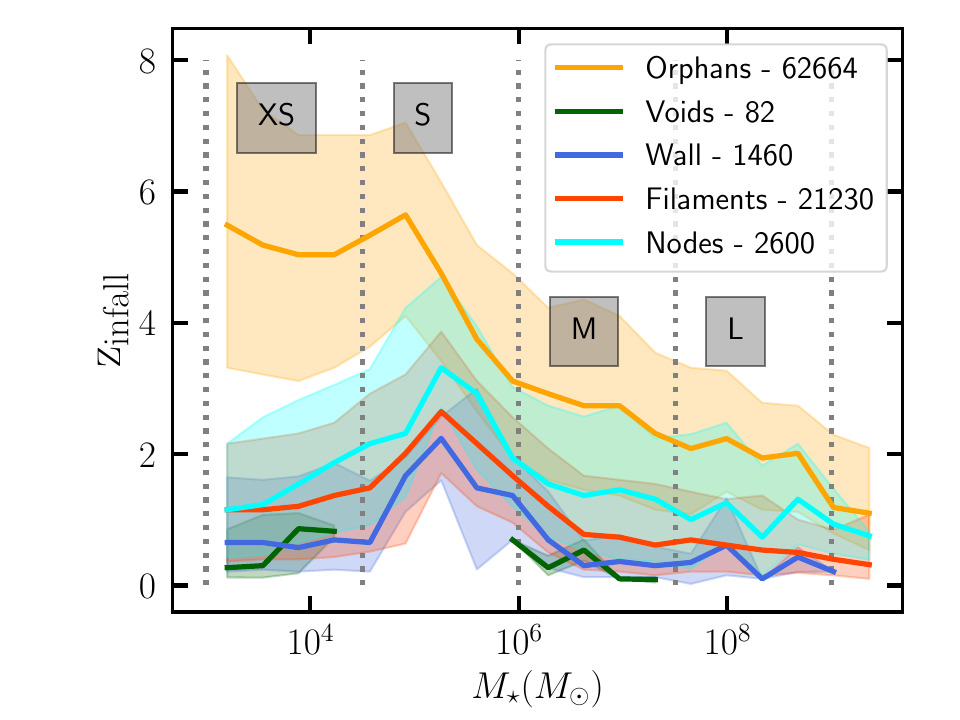}
    \caption{A plot of average redshift of infall vs. stellar mass. The lines show the 50th percentile of the infall time at that specific mass, and the shaded regions show the 16th to the 84th percentile infall times. The NEXUS+ environment groups contain only satellites, and the orphans are shown separately. Node galaxies infall on average earlier than other environments, with their 50th percentile infall time being at $z = 1.5$ when averaged over all masses, and void galaxies infall on average 5.2 Gyrs later, with the average infall time across all masses at $z = 0.4$. This large discrepancy between the average infall times of satellites in the different environments is the cause of the large environmental differences in stellar mass assembly seen in Figure~\ref{fig:nexusSMA}}
    \label{fig:infall}
\end{figure}

Satellite galaxy halo assembly shows smaller trends than the central galaxy halo assembly because, although haloes in denser environments have more opportunity to grow through mergers, and grow faster, they also (if they are satellites) infall into their host haloes significantly earlier than galaxy haloes in less dense environments (see Figure~\ref{fig:infall}). These two effects work against one another, reducing the overall trend when compared to central halo mass assembly.

Why do central galaxies not translate this large environmental difference in halo growth times into a large environmental difference in their stellar mass assemblies? This is simply because although their halo mass growth slows down at later times, resulting in an earlier $z_{50}$ for the haloes, there is no reason why their star formation would slow down. Star formation in GALFORM is proportional to the amount of cold gas in the disc, and in a large-enough central galaxy there is no reason why that gas supply would be cut off without quenching by AGN, supernova feedback or reionisation. So, the node galaxies continue to form stars at the same, or a higher, rate than before, resulting in minimal trends in the in-situ stellar mass assemblies between different environments.

Referring back to Figure~\ref{fig:nexusSMA}, this large environmental difference in halo mass assembly is the culprit for the very small environmental trends found in the stellar mass assembly of central galaxies. The node galaxies form more of their stars earlier than void galaxies, which catch up to them later resulting in them being similar masses at $z=0$. These trends are very small because the central node galaxies continue to form stars over time, reducing the effect of more early star formation on their 50\% and 90\% assembly times. However, this cannot explain the environmental trends in the stellar mass assembly of the satellite galaxies, because for them, the average $z_{50}$ difference in the largest three mass bins is 1.5 Gyrs, whereas for the haloes the average difference is 1 Gyrs. Therefore, we look to an aspect of halo differences which only affects satellites, which is the time at which the galaxy went from being a central to infalling into another halo and becoming a satellite.

Figure~\ref{fig:infall} explains the strong trends in environmental impact on stellar mass assembly found in satellite galaxies. Here, we see a histogram of the infall time of satellites and orphans in each NEXUS+ environment. The denser the large scale structure definition, the earlier galaxies infall and become satellites. 

Satellites in voids across all masses infall on average at $z = 0.4$, whereas the average infall time for satellites in nodes is at $z = 1.5$, 5.2 Gyrs earlier than satellites in voids. This means that the environment a galaxy resides in has a big impact on when a central becomes a satellite. This affects the stellar mass assembly because once the satellite is embedded in its host halo, it can no longer cool gas, and must use up whatever it has left: the infall starts the countdown to the end of their star formation. The difference in infall times between different regions of the cosmic web therefore drives the bulk of the differences we have observed for the stellar mass assembly of satellite galaxies.%Void satellites therefore have 5.2 Gyrs more (on average) in which to cool gas and form stars, and during the time in the universe of the highest star formation rate density \citep{madau2014}.}

This is also in agreement with what the findings in Evans et al. in prep, in which the different areas of large scale structure found in NEXUS+ show no differences in the number of mergers experienced, only in the average times when those mergers occurred.

\subsection{Altering the fiducial reionisation model}
\label{sec:alteration}
\begin{figure*}
	\includegraphics[width=\textwidth]{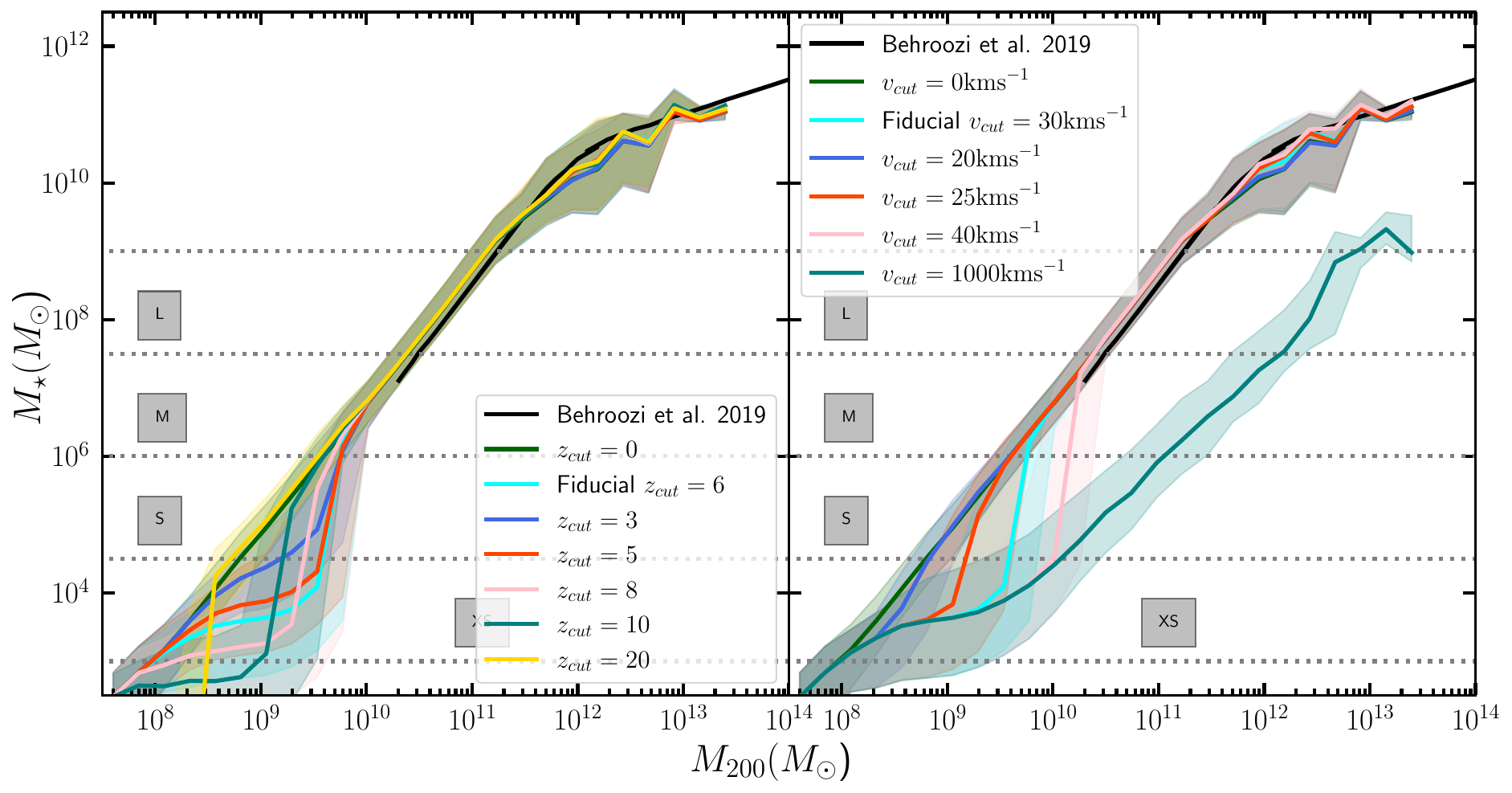}
    \caption{Stellar-halo mass relation of central galaxies for the different reionisation parameters. Different models are shown in different colours, with the 50th percentile of the stellar-halo mass relation shown by the solid lines, and the 16th to 84th percentile shown by the shaded regions. Two reionisation parameters were changed, the $z_{cut}$, which is the redshift at which reionisation happens, and the $v_{cut}$, which is the minimum circular velocity of the haloes which allow them to cool gas into them, post-reionisation. For the fiducial model, $z_{cut} = 6$ and $v_{cut} = 30{}$km{}~s$^{-1}$, and each model only changes one parameter from the fiducial model, i.e. all $v_{cut}$ models in the right panel have $z_{cut} = 6$. %From our most extreme models in the right panel, such as $v_{cut} = 1000$ and $v_{cut} = 0$, it is clear that the steep region of the stellar-to-halo-mass relation at roughly $M_{200} = 10^{9.7} $\Msun is created by the galaxies between those never affected by reionisation, and those always affected by reionisation, ie. the ``turnover dwarfs". Altering the reionisation parameters therefore changes the location and steepness of this region. For the left panel, we can see that changing the redshift at which reionisation occurs affects the depth of the difference between reionisation relics and normal dwarfs. For example, both $z_{cut}=0$ and $z_{cut}=100$ have no reionisation relics, because reionisation either happened from the formation of the first galaxies, or never occurred in the simulation. 
    Comparing the two panels, it is clear to see that changing $z_{cut}$ results in a change to the depth of the trough caused by reionisation relics, i.e. the relics have a smaller stellar mass if reionisation happens earlier, and changing $v_{cut}$ results in a change to the haloes that are impacted by reionisation; a higher $v_{cut}$ results in galaxies with a lower stellar mass at fixed halo mass.}
    \label{fig:mstars_mhalo_diffmod}
\end{figure*}

\begin{figure*}
	\includegraphics[width=\textwidth]{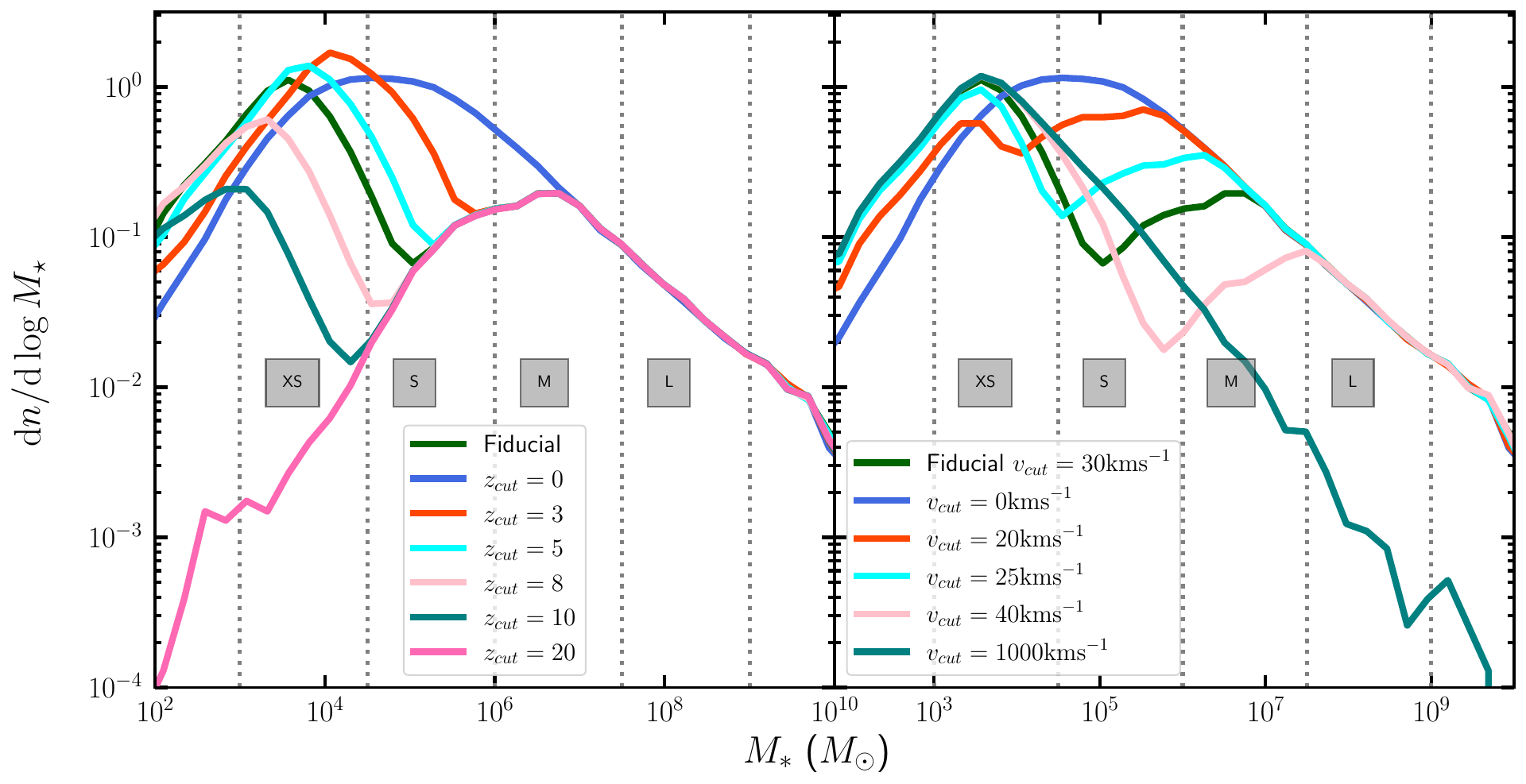}
    \caption{The stellar mass functions of all galaxy types predicted by the different models of reionisation. Different models are shown in different colours, with the left panel showing $z_{cut}$ variations and the right panel showing $v_{cut}$ variations. There is a characteristic dip between the reionisation relics and the regular dwarf galaxies across all models where reionisation is present, as shown also in Figure~\ref{fig:SMF_fiducial}. The $v_{cut}$ of the model affects the depth and location of the trough between the reionisation relics and galaxies unaffected by reionisation, specifically by changing how massive galaxies in the S and M mass bins are able to become. The greater the value of $v_{cut}$, the higher the masses of galaxies that have their growth suppressed by reionisation are. Changing $z_{cut}$ primarily changes the abundance of galaxies in the XS and S mass bins. The earlier reionisation occurs, the less time these relics have to build up mass before their star formation is cut off. In one of our extreme models, $z_{cut}=20$, reionisation relics are unable to form at all, and the only galaxies remaining are ones whose haloes have at some point been large enough to cool gas.}
    \label{fig:smf_alternate}
\end{figure*}

\begin{figure*}
	\includegraphics[width=\textwidth]{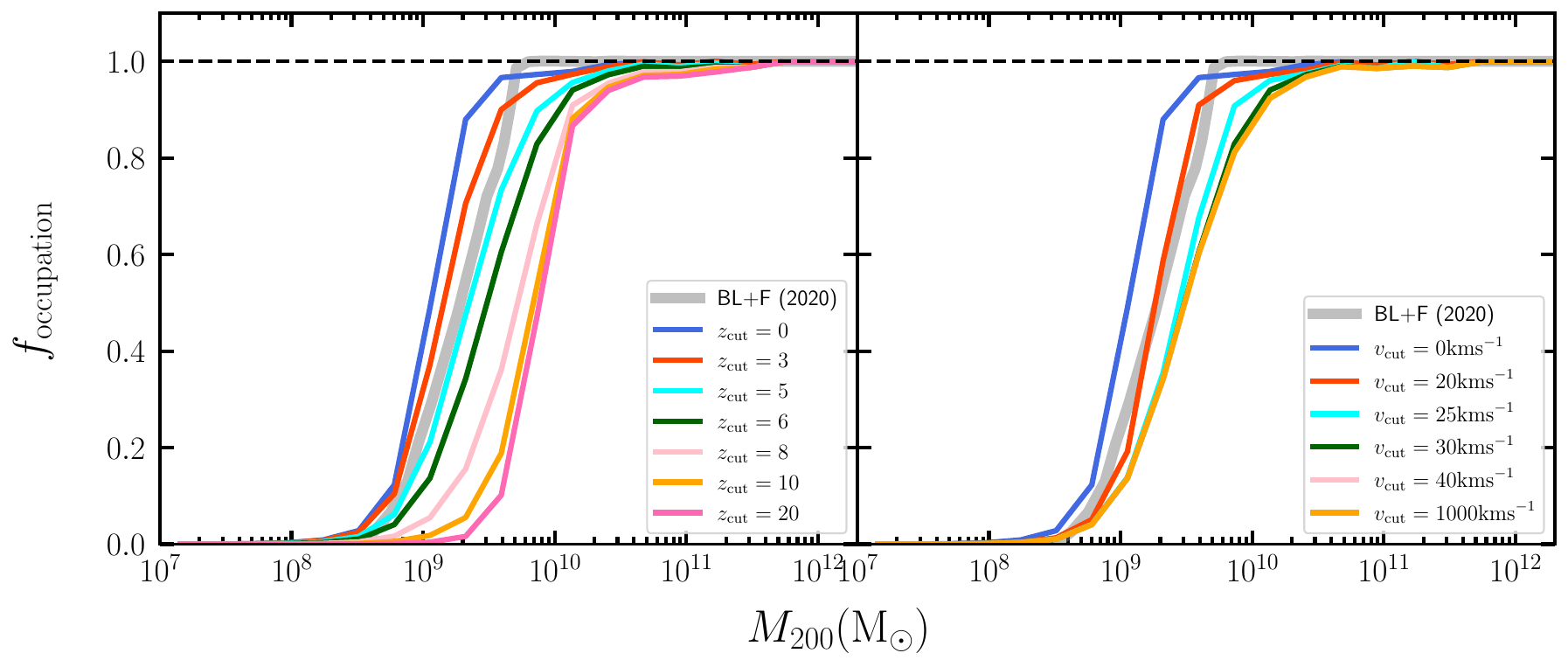}
    \caption{The halo occupation fraction for central galaxies at $z=0$ predicted by different reionisation models. Lower values of $v_{cut}$ and $z_{cut}$ result in higher occupation fractions at a given mass. The occupation fraction reaches a lower limit on the right panel, beyond which increasing $v_{cut}$ no longer has any effect at $v_{cut} = 40{}$km{}~s$^{-1}$, where the 50th percentile occupation occurs at a halo mass of $M_{200} = 10^{9.3} $\Msun. %This is because a set number of galaxies have already formed by $z = 6$, and the $v_{cut}$ only prevents galaxies from forming after $z = 6$. For the higher $v_{cut} = 1000 km s^{-1}$, the occupation fraction remains the same. 
    Changing $z_{cut}$ to an earlier time (higher redshift)  does converge towards a limit, which corresponds to the redshift before any galaxies form in the simulation, as seen in our most extreme model, $z_{cut}=20$.% Galaxies form in lower-mass haloes when reionisation is delayed.
    }
    \label{fig:occfrac_diffmod}
\end{figure*}

In this section, we explore the effect of altering the reionisation parameters on the environmental dependence of the stellar mass assembly of galaxies. As mentioned previously, GALFORM does not model reionisation in a spatially-dependent way, although it is well known in the literature that reionisation is patchy and extended in time (\citet{kannan2022, werre2025}). We will discuss this in more detail in Section~\ref{sec:discussion}. In the present exercise, we simply examine how changing parameters in our reionisation model manifest in different regions of the cosmic web. %, but the overall assumption is that reionisation happens at different times in different regions of the universe though, so by examining what a changed reionisation parameter would look like, we can approximate what effect it would have on environment.

When changing the parameters, we choose six additional variations on the redshift of reionisation, $z_{cut}$, and five additional variations on the circular velocity at which reionisation was said to kick in, or $v_{cut}$. In the fiducial model, $z_{cut} = 6$ and $v_{cut} = 30{}$km{}~s$^{-1}$. Figure~\ref{fig:mstars_mhalo_diffmod} shows the central galaxy stellar-to-halo mass relation for variations of the $z_{cut}$ parameter in the left panel, and for variations of the $v_{cut}$ parameter in the right panel. In the right panel, the effect of increasing $v_{cut}$ decreases the stellar masses of the galaxies below the mass defined by the circular velocity cut at $z\sim 0$. In our most extreme models, where reionisation either never affects any halo ($v_{cut} = 0{}$km{}~s$^{-1}$), or stops star formation in every halo ($v_{cut} = 1000{}$km{}~s$^{-1}$), the effects are clear. When reionisation never happens ($v_{cut} = 0{}$km{}~s$^{-1}$) we see that there is an unbroken power-law between the halo masses of galaxies and their stellar masses, where for each decade increase in halo mass, there is a 1.5 decade increase in stellar mass, until we reach the highest masses regulated by AGN feedback, at halo mass $M_{200} \geq 10^{12} $\Msun. On the other hand, when reionisation affects every halo ($v_{cut} = 1000{}$km{}~s$^{-1}$), we see a universal suppression in the stellar mass content of all haloes.%, in line with the suppression of star formation occurring for galaxies where $\Msun M_{\star} \leq 10^4 $\Msun. 
In this model, we are essentially seeing a snapshot of the stellar-halo mass relation at $z=6$, when reionisation kicks in. After this time, very few galaxies in this model grow.

Our other variations of $v_{cut}$ are bracketed by these extreme variations. In each case, there is essentially a change in the mass range at which the transition between the galaxies whose star formation is suppressed by reionisation at $z=6$ and those that are able to keep forming stars for longer. For $v_{cut} = 40{}$km{}~s$^{-1}$, the transition is steepest, with the most variation in stellar masses for the smallest range in halo masses, and for the $v_{cut} = 20{}$km{}~s$^{-1}$ model, the transition is more gentle, with barely any variation. Therefore, the presence and location of a ``kink''  in the true stellar-to-halo mass relation can help us to constrain the physics of reionisation \citep{bose2018}.

The effect of changing the $z_{cut}$ parameter is more subtle, but it is to largely change the halo mass range of the steepest portion of the relation, occupied by turnover dwarfs, with higher $z_{cut}$ values resulting in a lower halo mass for this region to be found in. When reionisation occurs at $z=10$, this steep region is centred around haloes of $M_{200}=10^{9.0}$\Msun, whereas if reionisation occurs at $z=5,6$, the steep region is centred on $M_{200}=10^{9.3}$ \Msun. This is because if reionisation occurs at $z = 10$, haloes will have less time to form stars before reionisation, lengthening the more horizontal region with galaxies less than $M_{\star} = 10^4$\Msun, and fewer haloes will have time to build up enough dark matter mass to cross the atomic cooling limit before reionisation. This also results in lower stellar masses across the entire halo mass spectrum, the shaded region showing the 16th-84th percentiles in Figure~\ref{fig:mstars_mhalo_diffmod} is much lower than other models for $z_{cut} = 10$. Simply put, the earlier the onset of reionisation, the longer and deeper the trough caused by suppressed star formation due to reionisation. The exception to this is our most extreme model, $z_{cut} = 20$, for which no relic galaxies form at all. More recent reionisation causes less suppression of star formation compared with earlier reionisation times. Our results here are therefore comparable to those in \citet{kim2024}, where the earlier the $z_{cut}$ of reionisation is, the more stellar mass is able to build up in this region of the SMHM relation, making it appear ``higher''.

The effect of changing reionisation parameters on the differential stellar mass function is stark - this is seen in Figure~\ref{fig:smf_alternate}. Changing the $z_{cut}$ parameter to its maximum value leads to a population of galaxies without any reionisation relics, and changing our $v_{cut}$ parameter to its maximum value leads to a population in which every galaxy is a reionisation relic. The characteristic dip is present in all models in which there are two populations: relics and ``normal'' galaxies, i.e. those not affected greatly by reionisation \citep{bose2018}. The masses of galaxies in the dip indicates the masses of the ``turnover'' dwarfs for each model. For all reasonable models of reionisation, this dip is found in the dwarf regime, at around $M_{\star} = 10^5 $\Msun. These are galaxies that are not reionisation relics, but also have not continued to grow in mass in the way that larger dwarfs have. As a result, this mass range is dominated by satellite and orphan galaxies, as well as late-forming central galaxies that will continue to grow into larger dwarfs. The turnover region shifts in accordance with the change in the stellar-to-halo-mass relation as $z_{cut}$ is changed (see Figure~\ref{fig:mstars_mhalo_diffmod}), even though the haloes in the turnover region are the same haloes as in the fiducial model, because $v_{cut}$ has not changed.

In Figure~\ref{fig:occfrac_diffmod}, we can see the effect on the occupation fraction of the central haloes in the different reionisation models. Increasing the value of $z_{cut}$ or the value of $v_{cut}$ causes more suppression of star formation, leading to a lower occupation fraction compared with other models in which $z_{cut}$ and $v_{cut}$ are lower. For example, in the left panel, we see that the halo mass at which the occupation fraction is 50\% is $M_{200} = 10^{9.1}$ \Msun{} for the model with no reionisation, $z_{cut} = 0$. For the model with the earliest reionisation, $z_{cut} = 20$, we see that 50\% occupation occurs at $M_{200} = 10^{9.9}$ \Msun. For the right panel, we see that for $v_{cut} = 0{}$km{}~s$^{-1}$, the occupation fraction is the same as for $z_{cut} = 0$, and for $v_{cut} = 1000{}$km{}~s$^{-1}$, the occupation fraction is 50\% at $M_{200} = 10^{9.5}$\Msun, which is the minimum occupation fraction possible when changing only the $v_{cut}$ parameter. The minimum occupation fraction achievable for the $v_{cut}$ variations is set by the haloes that were able to form stars before reionisation (at $z = 6$); conversely, the minimum for the $z_{cut}$ variations is bounded by the galaxies that do not continue to persist below the filtering scale once reionisation occurs, which can be seen in our $z_{cut} = 20$ model. The mass scale below which no galaxies form in the dark matter haloes also changes for different $z_{cut}$ values. In the $z_{cut}=20$ model, the minimum halo mass for galaxies to form is $M_{200} \sim 10^{9}$\Msun, and the minimum mass for the $z_{cut}=0$ model is $M_{200} \sim 10^{8}$\Msun. This is because the haloes in the models where reionisation is later have more time to build up enough dark matter to cool gas into their haloes at the much lower atomic cooling limit threshold before reionisation makes the limit much higher. Galaxies in models where reionisation is earlier consequently have less time to build up enough dark matter to reach the atomic cooling limit before reionisation, and therefore the minimum mass of haloes with galaxies in them is higher at $z=0$. For a visual representation of the halo growth compared to the reionisation and atomic cooling limits, see Figure~\ref{fig:RI_halogrowth}.

\begin{figure*}
	\includegraphics[width=\textwidth]{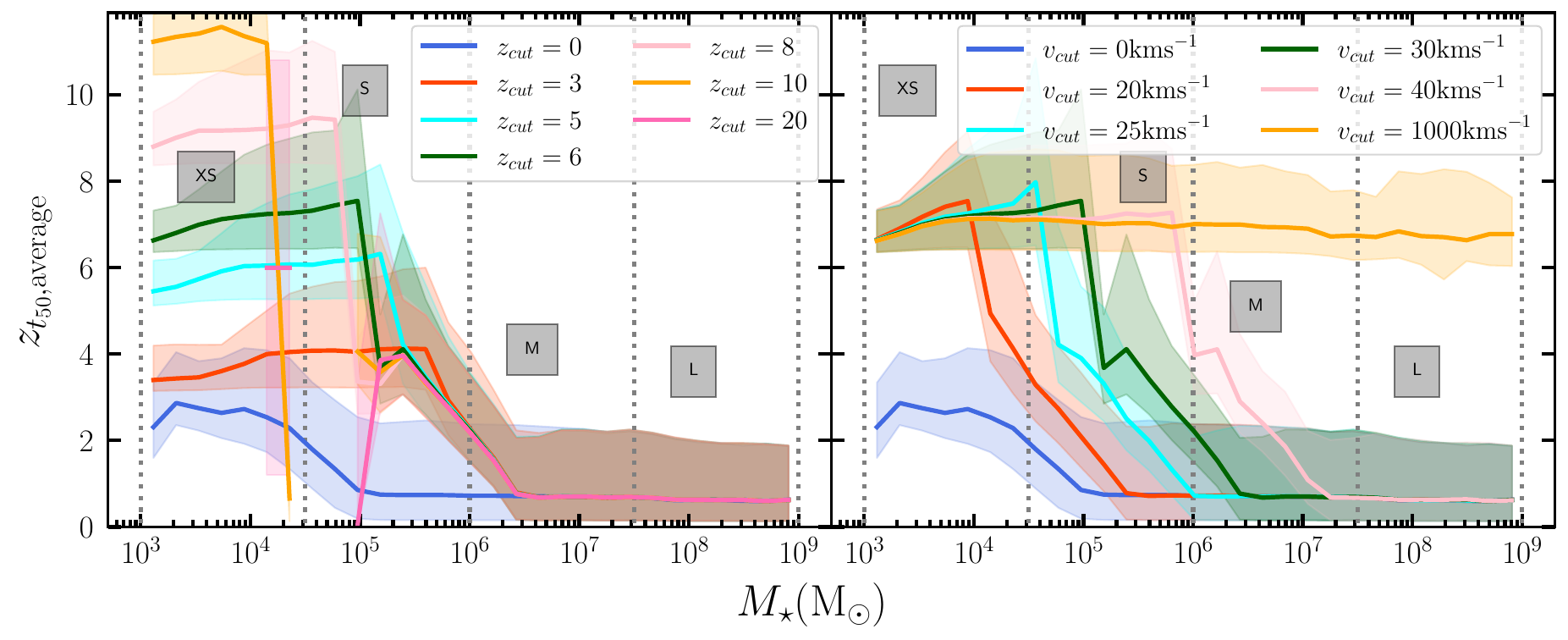}
    \caption{Plot of the average redshift at which central galaxies in a given reionisation model reach an in-situ stellar mass assembly of $50\%$. The shaded regions show the time at which the average galaxy reaches $10\%$ to $90\%$ of their assembly. For $v_{cut}$ variations, we can see a clear trend:  higher $v_{cut}$ values resulting in earlier assembly times, with more massive galaxies now becoming subject to the effects of reionisation. Changing the $z_{cut}$ parameter affects both the relic galaxies and the turnover galaxies of each model, but galaxies in the M and L mass bins are largely unaffected.
     }
    \label{fig:master_diffmod}
\end{figure*}

\begin{figure*}
	\includegraphics[width=\textwidth]{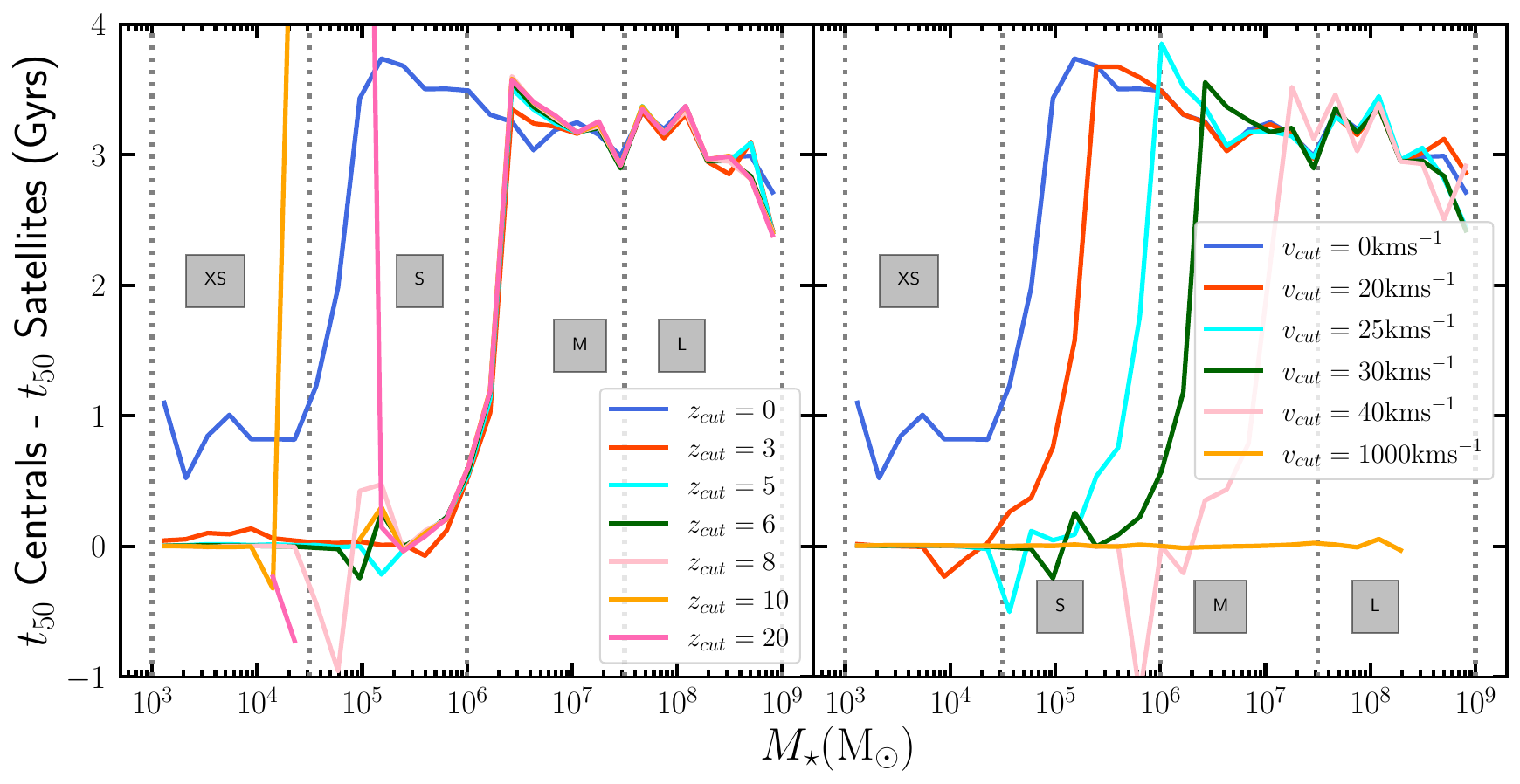}
    \caption{The difference between central and satellite galaxies for the 50\% stellar mass assembly time for the 50th percentile of galaxies. From this, we can see that, with the exception of the $z_{cut} = 0$ and $v_{cut} = 0{}$km{}~s$^{-1}$ models, all models have a transition between no differences in stellar mass assembly for very small galaxies and a difference of about 3 Gyrs for larger galaxies. For the $z_{cut}$ models, there is no difference for the mass ranges at which these differences occur, and they coincide with the location of the ``turnover region'' found in other figures. Whereas, for the $v_{cut}$ models, changing $v_{cut}$ changes the location of the turnover region, with a higher $v_{cut}$ value leading to a higher mass range for the turnover region.}
    \label{fig:tdiff_master}
\end{figure*}

Figure~\ref{fig:master_diffmod} compares the half-mass formation redshift ($z_{50}$) for central galaxies in each of the reionisation models. A clear pattern emerges regarding the location of the turnover region and the effect on relic galaxies. For the $z_{cut}$ variations, an earlier reionisation time (ie. a higher $z_{cut}$) results in a suppression of star formation at lower masses, denoted by an earlier average $z_{50}$ for the relic mass bin, XS. For $z_{cut} = 20$, there are no relic galaxies, and the galaxies below the critical mass of roughly $M_{\star} = 10^5 $\Msun{} are all very recently forming, indicating that they are not true relics, but are instead very young galaxies that may become larger dwarfs in the future. For our $z_{cut} = 10$ model, the relic galaxies have an average $z_{50}$ of $z=11.4$, much higher than our fiducial model's relic $z_{50}$ of $z=7.7$. All the average $z_{50}$ values of relics from different reionisation prescriptions form a pattern in which $z_{50,relic} \leq z_{cut}$. The turnover regions of the different $z_{cut}$ variations also change, with earlier reionisation resulting in a lower occupation of the turnover region, and a wider gulf between the relic galaxies and larger dwarfs, which is also visible in Figure~\ref{fig:smf_alternate}. This is due to reionisation limiting the galaxies that can become relic dwarfs to those who have already began their formation before the $z_{cut}$, and also making the resulting galaxies smaller due to their shortened time to create stars before reionisation. In turn, decreasing $z_{cut}$ results in the formation times of relic galaxies being more recent. In our $z_{cut} = 3$ model, the relic galaxies reach $z_{50}$ on average at $z = 4.2$, and this pattern is present across all $z_{cut}$ models, that the $z_{50}$ of relic galaxies will always be before the $z_{cut}$. In our $z_{cut} = 0$, which is equivalent to our $v_{cut} = 0{}$km{}~s$^{-1}$ model, in that reionisation never occurs, there are no reionisation relics. The smallest galaxies are instead quenched by their own supernova feedback, quenching later at $z \sim 3$.

In our $v_{cut}$ variations, the relationship between changing the parameter and the effect on the galaxies is more visually obvious. Increasing the $v_{cut}$ of the model leads to the turnover region moving to higher masses, and the relic region extending into the masses that would previously have been turnover dwarfs. For example, for the $v_{cut} = 40{}$km{}~s$^{-1}$ model, the highest mass galaxies within the turnover region are $M_{\star} = 10^{7.1} $\Msun, whereas in the $v_{cut} = 30{}$km{}~s$^{-1}$ fiducial model, the highest mass the turnover region extends to is $M_{\star} = 10^{6.2} $\Msun. The limits to this variation are our two most extreme $v_{cut}$ models, $v_{cut} = 0{}$km{}~s$^{-1}$ and $v_{cut} = 1000{}$km{}~s$^{-1}$. For $v_{cut} = 0{}$km{}~s$^{-1}$, reionisation never happens, equivalent to the $z_{cut} = 0$ model. It has a ``bump'' centred on $M_{\star} = 10^{3.75} $\Msun, which is associated with galaxies that have self-quenched at earlier times due to e.g., supernova feedback. For $v_{cut} = 1000{}$km{}~s$^{-1}$, reionisation affects every halo at $z = 6$, creating a population of only reionisation relic galaxies, which quench at $z \sim 7.5$. Between these two values, the reionisation parameters being adjusted just changes the mass of the turnover region between these two groups.

In Figure~\ref{fig:tdiff_master}, the differences between the different models and their satellite and central galaxy assembly times are shown. Here, we take the difference between the respective $z_{50}$ values for centrals and satellites in each model. The satellites across both kinds of reionisation parameter adjustments show a similar pattern of delayed assembly, with the satellites assembling on average 3 Gyrs earlier than the central galaxies for almost all models and for non-relic masses. For the lowest mass galaxies, our ``relics'', we find no difference between satellite and central galaxies, because these galaxies are quenched by reionisation well before the average satellite infall time. The $z_{cut} = 0$ and $v_{cut} = 0{}$km{}~s$^{-1}$ models, which are essentially the same in that no galaxies are reionised, also have smaller differences between satellite and central galaxy assembly at lower masses, due to quenching by supernova feedback. The turnover region for the $z_{cut}$ models is at $M_{\star} = 10^5 - 10^6 $\Msun, and higher than this, the galaxies assemble with a 3 Gyr delay on average, with this decreasing slightly at higher masses. This decrease is due to the fact that larger satellite galaxies are biased towards a later infall time, because they need more time to build up stars before infalling. This is also what we expect from hierarchical structure formation models like $\Lambda$CDM: larger dark matter haloes infall later.

%This is why you would expect to see the differences in assembly in the dwarf regime, and only smaller differences in much larger galaxies.}

For the different $v_{cut}$ models in Figure~\ref{fig:tdiff_master}, the turnover region changes, with lower values of $v_{cut}$ associated with lower-mass turnover regions and higher values of $v_{cut}$ associated with higher-mass turnover regions. For example, for $v_{cut} = 20{}$km{}~s$^{-1}$, the turnover region is between $M_{\star} = 10^{3.5} - 10^{5} $\Msun, whereas for the $v_{cut} = 40{}$km{}~s$^{-1}$ model, the turnover region is between $M_{\star} = 10^5 - 10^7 $\Msun.

\section{Discussion}
\label{sec:discussion}
%\WH{Add one short diagnostic sentence to make the causal interpretation of `satellite infall bias' more robust: propose an infall-time-matched null test. Even if we do not run it in this draft, stating it as a clear check strengthens the argument and preempts referee pushback.} \Mac{Yep, agreed this is a good idea.}

The main result in this paper is the influence of ``satellite infall bias” on the differences in assembly histories between satellite galaxies of the same final stellar mass between different parts of the cosmic web, compared with the more similar assembly histories of central galaxies. Satellite infall bias refers to environmental differences in satellite properties caused by variations in their typical infall times. The 50th percentile satellite infall time is $z = 1.5$ for node galaxies, and $z = 0.4$ for void galaxies (see Figure~\ref{fig:infall}), which is a difference of 5.2 Gyrs. This results in assembly history $z_{50}$ differences between these regions for satellite galaxies of the same mass, but only in mass bins where the growth of the galaxies is not immediately suppressed by reionisation at $z=6$.
A direct diagnostic of this mechanism is to compare satellites across environments at fixed final stellar mass after matching their infall-time distributions; if infall-time bias is the dominant driver, the residual differences in $z_{50}$ between environments should be substantially reduced under such a matching.

The \citep{lacey2016} GALFORM model assumes an instant ram-pressure stripping of the hot gas haloes from satellite galaxies. This is an assumption made for simplicity, and in reality, the stripping would not be instantaneous. \citet{wetzel2013} finds that satellite galaxies should continue to form stars as normal for 2-4 Gyrs after infall, at which point, they then rapidly quench. This is also consistent with hydrodynamic and magnetohydrodynamic simulations of satellite infall which predict that the hot haloes of galaxies take 2-4 Gyrs to quench \citep{rintoul2025}. This was a discrepancy that the \citet{hou2018} GALFORM model attempted to fix, employing a different gas cooling model and succeeding in producing “bluer” satellites, more consistent with observations.
%\WH{This sentence over-claims: given modelling assumptions (instant stripping in fiducial GALFORM, no spatially dependent reionisation, etc.), please avoid `we should already see'. Rephrase as a conditional, testable statement.} \Mac{Yep, thanks for the rephrase, I agree it's clearer.}
%\sout{However, considering a 2-4 Gyr delay for hot halo loss with our results, even considering environmental effects not implemented in GALFORM such as more rapid satellite stripping in nodes, we should already see the differences in stellar mass assembly in the real universe.
However, even factoring in a 2--4~Gyr delay for hot-halo loss, the large separation in typical infall times between node and void satellites implies that the corresponding differences in stellar-mass assembly should be testable observationally, at least statistically, once sufficiently complete dwarf samples and robust environment metrics become available.
% Wojciech rephrased above statement
The delay in satellite infall for voids compared to nodes is 5.2 Gyrs, and the time between average void satellite infall ($z=0.4$) and $z=0$ is 4 Gyrs, meaning that the 50th percentile satellites in these regions are already quenched. If denser environments quench satellite galaxies faster, these differences in stellar mass assembly between void satellites and node satellites will be even greater than found in GALFORM.

Central galaxies in COCO/GALFORM do not show the same differences in $z_{50}$ across the cosmic web, despite the fact that the more rapid halo growth in denser regions should be expected to produce differences in assembly history. From Figure~\ref{fig:nexusSMA}, in the M mass bin, at $z=6$ the node galaxies were significantly larger than galaxies in other regions, allowing them to continue to form stars while the other regions faced a brief pause in star formation.  By $z \sim 3$, the other regions caught up to the node region central galaxies while the node galaxy growth slowed, due to the node galaxy haloes for this mass bin slowing in growth (see Figure~\ref{fig:nexus_halo_SMA}). At that time, the majority of mergers had already occurred in node regions, leading to them running out of material and dark matter to accrete. By $z=0$, node galaxy centrals are in significantly smaller haloes than galaxies of a similar stellar mass in other regions, and although their stellar mass assembly history curve has a different shape to other regions, the times at which they reach $z_{50}$ and $z_{90}$ are the same. This is partially a result of serendipity due to the length of time the universe has evolved for, and if we evolved those galaxies for a further 10bn years, the $z_{90}$ and $z_{50}$ times would be more significantly different. The other cause is simply that the differences in early stellar mass assembly are undone quickly by the effects of the smaller node galaxy haloes and their mergers all being completed earlier.

In reality, areas of higher density of galaxies will have different in-situ stellar mass assembly histories, even if they are central galaxies, as shown in papers observing the quenching of central galaxies in clusters \citep{wetzel2014}. In GALFORM, environmental effects are encoded as results from mergers or changes in galaxy type from central to satellite. As such, it is useful to compare to results from hydrodynamic simulations. \citet{christiansen2024} compares simulations of dwarf galaxies formed near Milky Way-like hosts with dwarf galaxies formed in isolated but filament-like structures. They find that more isolated dwarfs have lower stellar masses for their halo masses when compared to dwarf galaxies formed near a MW mass host, even before they infall as satellites. The dwarf galaxies in the higher-density environments also mostly formed their stars faster and earlier, finishing their halo and stellar mass assemblies before the more isolated dwarfs, and this was found for both satellite and central dwarf galaxies. These results are consistent with what we find in COCO/GALFORM, with the exception of the central galaxies’ stellar mass assemblies also being affected. \citet{christiansen2024} attributes this to potential dwarf-dwarf interactions or differences in timescale of gravitational collapse. Given that we don’t see this result in GALFORM, it is likely that the timescales of gravitational collapse are having a smaller effect than other environmental effects seen in their hydrodynamic simulation, as these timescales are different in GALFORM too.

Results from recent research using JWST \citep{kashino2023}, and from recent simulations \citep{dawoodbhoy2023}, indicate that reionisation was inhomogeneous and not instantaneous, as it is in GALFORM. Investigating the effect this bias could have on our results was the purpose of our experimenting with changing the reionisation parameters of GALFORM. The differences in assembly time between satellites and central galaxies in our different models of reionisation were invariant to changes in $z_{cut}$, and only varied in the turnover region when altering $v_{cut}$. If we accept that denser areas of the cosmic web are likely to reionise first, central node galaxies would have earlier stellar mass assemblies than central void galaxies. This would also lower the final stellar masses of node galaxies, resulting in the differences in the stellar-to-halo-mass relation between cosmic regions potentially disappearing (see right panel of Figure~\ref{fig:nexusSMF}).

Finally, the turnover regions caused by reionisation are a particularly interesting regime for future investigation. In our model, not only does this region represent a dip in the stellar mass function, but also a change in the types of galaxies found there. GALFORM labels disrupted subhaloes as ``orphan'' galaxies, meaning that they are below the halo resolution limit we have imposed, but we can continue to track their most bound particle. The turnover region represents the peak in the orphan fraction as a whole, caused by two factors: orphans have an earlier infall than regular satellites in our model, so we are really looking at very early infalling satellites in the real universe, and because there is a lack of central galaxies at this mass range. Specifically looking at galaxies in the trough of the stellar mass function, we find no central galaxies with these stellar masses, they are all satellites and orphans. This is due to them having fallen into their haloes roughly around the time of reionisation: a central galaxy of that mass at reionisation would have continued to grow, and a central galaxy of a slightly smaller mass would have stopped growing due to reionisation, becoming a reionisation relic galaxy. In the real universe, locating the turnover region caused by reionisation could be done by both constructing a stellar mass function from observed galaxies, and also locating the mass range where the proportion of central galaxies is lowest. Due to inhomogeneous reionisation, the turnover region would be shallower and encompass a wider range of stellar masses. Finding this turnover region would require an improvement in our capability to detect ultra-faint dwarf galaxies which may come with the opening of telescopes like the LSST, which had first light in June 2025.

\section{Conclusions}
The stellar mass assembly of dwarfs is the complex outcome of the physics of galaxy formation and the large-scale environment in which they grow. In particular, the differences in the assembly histories of dwarfs based on their specific location in the cosmic web is relatively unexplored in cosmological simulations. This is due to the difficulty of  resolving the smallest galaxies within a representative cosmological setting, as this requires very high resolution and an appreciably large volume making such an exploration computationally expensive. To overcome these limitations, in this work we make use of the {\it Copernicus Complexio} simulation, to which we apply the Durham semi-analytic model of galaxy formation, GALFORM, to create a representative population of galaxies spanning the entire mass range relevant to dwarfs in a cosmological volume. We divided the dwarfs into mass groups based on their final-day stellar mass, and we then assessed the impact of various different properties (mass, galaxy type, large-scale environment, local dark matter halo density) on their assembly histories. Furthermore, we varied the parameters that define the reionisation model in GALFORM to study the effects of reionisation on the final-day properties of these dwarfs. Our main results are summarised below:
\begin{enumerate}
    \item We found that in our sample, $z=0$ stellar mass had the largest impact on the stellar mass assembly, especially when considering smaller dwarf galaxies, more likely to be impacted by reionisation (Figure~\ref{fig:mass_SMA}). The largest stellar mass bin, ``L'' = $10^{7.5} - 10^9 $\Msun, assembled 50\% of its mass 7.7 Gyrs later than our smallest stellar mass bin, ``XS'' = $10^3 - 10^{4.5} $\Msun.
    \item Our first environmental test was based on the position of the galaxy in its local environment, and used the galaxy `types' contained within GALFORM, namely satellites, centrals, and orphans (Figure~\ref{fig:type_sma}). Between these groups, we found trends not only in the stellar mass assembly histories, but also that these trends were strongest in dwarf galaxies that were larger than reionisation relics ($M_{\star} > 10^5 $\Msun). This is because these satellites are able to keep evolving up to their infall time, which is later, on average, than $z = 6$, when reionisation begins in our fiducial model. The environmental differences in satellite infall time then are able to affect the stellar mass assemblies of satellites in different environments. Discounting our smallest mass bin, XS, satellite galaxies assemble 50\% of their mass an average of 4.3 Gyrs earlier than central galaxies. 
    \item Our second environmental test involved the application of the program NEXUS+ to the $z=0$ output of COCO/GALFORM (Figure~\ref{fig:nexusSMA}). This allowed us to distinguish between the large scale structure environments and compare between the different mass scales. We found that there were small dependencies on LSS environment for central galaxies, and larger discrepancies across most mass ranges for the satellite galaxies, the exception for this being the XS mass bin, because the average infall time of the satellites was after reionisation. The average difference between the densest and least-dense environments for assembling 50\% of their mass was 1.5 Gyrs for satellites and 0.08 Gyrs for central galaxies, excluding our XS mass bin.
    \item Our third environmental measure was using a measure of local environmental density, which measured the local environment as either in the $95$th percentile, $40-60$th percentile, or $5$th percentile, based on the number of dark matter haloes within $2$Mpc and outside of $0.2$Mpc (Figure~\ref{fig:tilly_sma}). These also showed trends in assembly similar to the NEXUS+ environment measure, with satellites in the least-dense environments assembling 50\% of their mass on average 1.4 Gyrs later than satellites in the densest environments, and central galaxies only showing a 0.37 Gyr difference, but trending the opposite way with denser environments finishing their assembly later. This also excludes our smallest mass bin, XS.
    \item The trends seen in central galaxies for the NEXUS+ and local density environment measures can be explained by the biases in the halo mass assembly in dense regions compared with sparse regions. These trends are much smaller than the trends in halo mass assembly because the haloes continue to form stars despite the halo mass plateauing, so long as the circular velocity of the halo is above the $v_{cut}$ defined in the model (Figure~\ref{fig:RI_halogrowth}). At higher masses, the stellar mass assemblies of galaxies forming earlier would be regulated by AGN feedback earlier, but this is not true in the dwarf regime in GALFORM, and so the galaxies continue to form stars as before.
    \item The greatest trends for both NEXUS+ environment and local density environment for stellar mass assembly were found in satellite galaxies (Figure~\ref{fig:nexus_halo_SMA}, Figure~\ref{fig:type_sma}). These trends can be explained by the differences in average infall times across different environments (Figure~\ref{fig:infall}). In dense environments, galaxies begin infalling into their host haloes earlier than in sparse environments, where it takes longer for galaxies to interact with one anther, resulting in later infall times of satellites. The average time difference between the median infall times of satellites in nodes and voids was 5.2 Gyr.
    \item Galaxy infall time is also the explanation for the unusual mix of galaxy types in the ``turnover'' stellar mass bin, S = $10^{4.5} - 10^6 $\Msun{} (Figure~\ref{fig:galtype_frac}). The mass range between reionisation relics and larger dwarfs has a far smaller proportion of central galaxies, reaching $\sim$ 0\% centrals, $\sim$ 80\% orphan galaxies, and $\sim$ 20\% satellite galaxies roughly at the same mass range as the dip in the stellar mass function. This is due to the fact that the satellite and orphan galaxies in this mass bin have an infall time very close to $z_{cut} = 6$, our fiducial reionisation time, and so their stellar masses are frozen in this unusual mass range. Central galaxies in this mass range at $z = 6$ are either large enough to form stars for longer, and so move up in mass, or so small that they become reionisation relics. As such, they are sorted out of this mass range, and the only galaxies left are orphans and satellites with very early infalls. Because void regions have delayed satellite infall, this region of the stellar mass function is therefore unpopulated in voids (Figure~\ref{fig:nexusSMF}).
    \item When changing our reionisation parameters, $z_{cut}$ and $v_{cut}$, we found that changing the timing of reionisation regulated the abundance of reionisation relic galaxies, and changing the strength of reionisation regulated the growth of the non-relic galaxies (Figure~\ref{fig:smf_alternate}). Earlier reionisation suppressed the relic population, and stronger reionisation suppressed the non-relic population.
    \item The effects of changing these parameters on the differences between satellite and central stellar mass assembly are found in Figure~\ref{fig:tdiff_master}. For more massive dwarfs, generally there is a 3 Gyr difference between the time at which central and satellite galaxies reach 50\% assembly, with satellite galaxies being first. This difference is non-existent for relic galaxies unless we set the timing of reionisation to be later than the average satellite infall, after $z=2$. Changing the $v_{cut}$ parameter shifts the mass range at which reionisation affects the stellar mass assemblies, and so a stronger reionisation results in more massive galaxies being relics, and showing no difference in their stellar mass assemblies for central and satellite galaxies. A weaker reionisation has the opposite effect, resulting in lower-mass galaxies showing differences between central and satellite assembly.
\end{enumerate}
%\WH{Conclusions/Outlook: the current outlook focuses only on LSST. Suggest expanding slightly to also mention 4HS/4MOST as a near-term, highly complementary spectroscopy+environment data set for $z\lesssim 0.15$, and connect to COCO-based predictions for very faint/low-surface-brightness structures (stellar haloes/streams and ghost-galaxy scenarios) that can be tested with deep imaging. Keep the claims conditional and avoid overpromising resolved SFHs at large distances.} \Mac{Thanks, I don't have great knowledge on upcoming obs surveys, so this is very helpful.}
As surveys like LSST begin releasing their data in the next few years, we will expect to see a sudden expansion in the amount of data we have on the smallest dwarf galaxies. While it will still be challenging to obtain resolved-star formation histories for large samples beyond the very local volume, the greatly improved census of dwarf populations will provide important tests of the predicted mass dependence of reionisation effects and of satellite versus central assembly trends. In parallel, upcoming wide-area spectroscopy such as the 4MOST Hemisphere Survey (4HS) will deliver highly complete redshift samples out to $z\lesssim 0.15$, enabling robust environmental characterisation for millions of galaxies and offering a natural route to test statistical predictions for environment-dependent dwarf growth \citep{Taylor2023_4HS}. Finally, deep low-surface-brightness imaging and stellar-stream searches will increasingly allow direct tests of COCO-based predictions for accreted stellar components and tidal debris around low-mass galaxies \citep[e.g.][]{Cooper2025Accreted,Deason2022DwarfHalos,MiroCarretero2025Streams}, as well as rare accretion-dominated `ghost galaxy' scenarios \citep{Wang2023Ghostly}.
% Above section rephrased by Wojciech
Our conclusions rely on approximations inherent to the semi-analytic approach, which can be tested and refined in future work guided by next-generation hydrodynamical codes as well as upcoming data on dwarf galaxies.
% above section suggested by Isabel
The next step forward for our work, and similar conclusions on the impact of reionisation on dwarf galaxies, is to compare our results to cosmological simulations with physically-motivated reionisation models implemented (e.g. patchy reionisation), whether these are semi-analytic implementations or using radiation hydrodynamics simulations.
% this above phrase is new from Mac, and I don't know how helpful it is, but it does seem like it would be the next step to verify these kinds of results.

%\WHED{\sout{As surveys like LSST begin releasing their data in the next few years, we will expect to see a sudden huge expansion in the amount of data we have on the smallest dwarf galaxies. As this comes about, it is still unlikely we will be able to compare the impacts of large scale structure on dwarf galaxy stellar mass assembly, due to the resolution being inadequate for fully resolving stellar populations out to that distance. However, this may help affirm the usefulness of our models for higher mass dwarfs, and may lead to exciting new discoveries in the baryonic processes happening inside dwarf galaxies, which will be useful to update our model with. Our conclusions also rely on approximations that can be tested with other hydrodynamic simulations, and we look forward to more investigations from theorists to tell us if we can expect to see our results replicated in the real universe.}}

\section*{Acknowledgements}
We thank Krishna Naidoo and the \texttt{CACTUS} development team for support regarding the \texttt{CACTUS} implementation of the NEXUS+ cosmic-web classification used in this work ~\citep{Hunde2025}. SB is supported by the UKRI Future Leaders Fellowship [grant numbers MR/V023381/1 and UKRI2044]. AF acknowledges support by a UK Research and Innovation (UKRI) Future Leaders Fellowship [grant no MR/T042362/1] and a Sweden's Wallenberg Academy Fellowship. ISS acknowledges support from the European Research Council (ERC) Advanced Investigator grant to C.S. Frenk, DMIDAS (GA 786910) and from the Science and Technology Facilities Council [ST/P000541/1] and [ST/X001075/1]. This work used the DiRAC@Durham facility managed by the Institute for Computational Cosmology on behalf of the STFC DiRAC HPC Facility (www.dirac.ac.uk). The equipment was funded by BEIS capital funding via STFC capital grants ST/K00042X/1, ST/P002293/1, ST/R002371/1 and ST/S002502/1, Durham University and STFC operations grant ST/R000832/1. DiRAC is part of the National e-Infrastructure.

%%%%%%%%%%%%%%%%%%%%%%%%%%%%%%%%%%%%%%%%%%%%%%%%%%
\section*{Data Availability}

The data used in this paper can be made available upon request to the corresponding author. 



\bibliographystyle{mnras}
\bibliography{coco_environment} % if your bibtex file is called example.bib

%\begin{thebibliography}{99}
%\end{thebibliography}

%%%%%%%%%%%%%%%%%%%%%%%%%%%%%%%%%%%%%%%%%%%%%%%%%%

%%%%%%%%%%%%%%%%% APPENDICES %%%%%%%%%%%%%%%%%%%%%

% \appendix

% \section{Some extra material}

% If you want to present additional material which would interrupt the flow of the main paper,
% it can be placed in an Appendix which appears after the list of references.

%%%%%%%%%%%%%%%%%%%%%%%%%%%%%%%%%%%%%%%%%%%%%%%%%%

% Don't change these lines
\bsp	% typesetting comment
\label{lastpage}
\end{document}